\documentclass[aps,pre,twocolumn,superscriptaddress]{revtex4-1}
\usepackage{graphicx,amsmath,epsfig,psfrag,verbatim,color,enumitem,float}
\usepackage[]{xcolor}
\usepackage[caption=false]{subfig} 
\usepackage[normalem]{ulem}
\usepackage{epstopdf}
\usepackage{comment}
\usepackage[normalem]{ulem}  

\definecolor{darkgreen}{rgb}{0,0.6,0} 

\begin{document}
\title{\textcolor{black}{Emergent Time Crystal} with Tunable Period \\
in the Uniaxial Random Field XY Magnet}

\author{S. Basak}
\affiliation{Department of Physics and Astronomy, Purdue University, West Lafayette, IN 47907, USA}
\affiliation{Purdue Quantum Science and Engineering Institute, West Lafayette, IN 47907, USA}
\author{K.~A. Dahmen}
\affiliation{Department of Physics, University of Illinois at Urbana-Champaign, Urbana, IL 61801, USA}
\author{E.~W. Carlson}
\email{ewcarlson@purdue.edu}
\affiliation{Department of Physics and Astronomy, Purdue University, West Lafayette, IN 47907, USA}
\affiliation{Purdue Quantum Science and Engineering Institute, West Lafayette, IN 47907, USA}

\date{\today}

\begin{abstract}
The addition of uniaxial random fields to the XY model induces an order-by disorder transition,
in which the XY magnet develops a spontaneous magnetization in the direction
perpendicular to the uniaxial random field.  
Here, we use simulations to explore the robustness of this phase transition with respect to a rotating driving field.
We find that the order-by-disorder transition is robust, persisting to finite applied field.
In the vicinity of the critical driving field strength, a time crystal emerges, 
in which the period of the limit cycles becomes an integer $n>1$ multiple
of the driving period.  Because $n$ increases with system size, 
the period of the time crystal can be engineered.
This period multiplication cascade is reminiscent of that occuring in
amorphous solids subject to oscillatory shear near the onset of plastic deformation,
and of the period bifurcation cascade near the onset of chaos in nonlinear systems,
suggesting it is part of a larger class of phenomena in transitions of dynamical systems.
Applications include magnets, electron nematics, and quantum gases.
\end{abstract}

\maketitle

\section{Introduction}
\label{sxn:Introduction}
The XY model, 
in which interacting spins are confined to rotate within a plane,
has been a staple of statistical mechanics and condensed matter studies, 
having been applied to a broad range of physical systems including
planar magnets, superfluids, superconductors, two-dimensional melting, 
nematic liquid crystals, and electron nematics, among others.  
\cite{lacour-gayet_toulouse,PhysRevB.73.060504,PhysRevB.74.224448,PhysRevB.23.335,PhysRevB.19.2457,PhysRevLett.41.121,PhysRevB.19.1855,Lilly-1999-PhysRevLett.82.394,abanin-lee,girvin-macdonald-book}
In two dimensions, the XY model 
exhibits a BKT transition  to a power-law ordered phase, yet
with no long-range order.\cite{berezinskii,kosterlitz-thouless}  
As such, the addition of random fields to a two-dimensional XY model is expected
to result in even less order:  Imry and Ma argued that a ($d \le 4$)-dimensional system with continuous order parameter (with O(n) symmetry with $n \ge 2$) in the presence of random fields cannot have long range order for any finite disorder strength.~\cite{PhysRevLett.35.1399}

However, the addition of {\em uniaxial} random fields reduces the global symmetry of the
Hamiltonian, and the Imry-Ma argument no longer applies.\cite{Crawford2013}  
In this case, the low temperature phase has long-range order via an order-by-disorder transition,
in which XY spins align perpendicular to the random fields.\cite{Crawford2013,PhysRevB.32.3081}
This is a special case of a more general class of order-by-disorder transition, where an n-dimensional spin system orders in a  (n-k)-dimensional subspace due to orthogonal k-dimensional random fields.~\cite{Crawford2013,Feldman1999,PhysRevB.74.224448,abanin-lee}

In this paper, we consider the possibility of a nonequilibrium transition.
We use simulations to study the order-by-disorder transition in the presence of 
a rotating driving field.  
By analyzing the avalanche size distribution as a function of magnitude of applied driving field. 
We find evidence that the system undergoes a continuous nonequilibrium
phase transition at a critical amplitude of the driving field.
Once a limit cycle is established, 
we observe that the period of the hysteresis loops become n-fold 
near a critical applied field strength, where n is as large as 7 in our largest systems.  
We present evidence that the period of the subharmonic entrainment
is rigid against perturbations in initial conditions, and against perturbations of the drive
field, indicating that a classical discrete time crystal emerges at criticality.\cite{yao-nayak-2018,wilczek_cdtc_2012}
We present finite size scaling evidence that 
the period of these multi-period limit cycles will diverge in the thermodynamic limit. 
An experimental test of this would be the presence of 
non-repeatability in the response due to a
rotating driving field near the transition.

As discussed further in Sec.~\ref{sxn:applications}, 
there are several experimental systems 
corresponding to the XY model into which
uniaxial random field disorder can be incorporated, 
whereby these ideas can be tested experimentally.
These include
layers of Josephson junctions,\cite{lacour-gayet_toulouse}
superfluid in a uniaxially stressed aerogel,\cite{PhysRevB.73.060504} 
ultracold atoms in the presence of speckle radiation,\cite{PhysRevB.74.224448}
uniaxially stressed 2D Wigner crystals,\cite{PhysRevB.23.335,PhysRevB.19.2457,PhysRevLett.41.121,PhysRevB.19.1855} 
the half-integer quantum Hall effect,\cite{Lilly-1999-PhysRevLett.82.394} 
and possibly the graphene quantum Hall ferromagnet.~\cite{abanin-lee,girvin-macdonald-book}

\section{Model}
\label{sxn:2D_XY_model_uniaxial_RF}

We consider the uniaxial random field XY model on a square lattice,
in the presence of a driving applied field $H$: 
\begin{eqnarray}
  \mathcal{H}=-J\sum_{\langle i,j \rangle}\cos{(\theta_{i}-\theta_{j})} 
  &-&\sum_{i}h_i\cos{(\theta_{i})}  \nonumber \\
  &-&H\sum_{i}\cos{(\theta_{i}-\phi)},
  \label{eqn:model_XY_uniax_RF}
\end{eqnarray}
where $\vec{S}_i \equiv ( \cos(\theta_i), \sin(\theta_i) ) $ 
is the XY spin on each site $i$, 
and $J$ is the nearest neighbor interaction strength.
The second term arises from the interaction of a local
random field along the x-axis and the XY spins.  
We choose the random field $h_i$ at each site $i$  from a gaussian probability distribution 
of width $R_x$, $P(h_i) = exp[-h_i^2/(2 R_x^2)]/(\sqrt{2 \pi R_x^2})$.  
The order parameter is the magnetization per site $\vec{m}=\frac{1}{N}\sum_{i=1}^{N} \vec{S}_i$, where 
$N = L \times L$ is the number of sites.

We study this system under the influence of a rotating applied driving field
whose angle $\phi = \Omega t$ advances in time slowly compared to
all other timescales in the problem.  
The dynamics is quasi-static:  after each small increment of the driving field angle,
the energy of the system is minimized.  (See Sec.~\ref{sec:methods} for 
details of the simulation method.)
This type of dynamics\cite{reichhardt-dahmen} presupposes that
the system is connected to a heat bath which prevents heating by the drive.

\begin{figure*}[htbp]
  \begin{center}
    \subfloat{%
     \label{sfig:4panel_Mx_My_for_h_160_multiplot}%
     \resizebox{1.0\textwidth}{!}{%
         \input{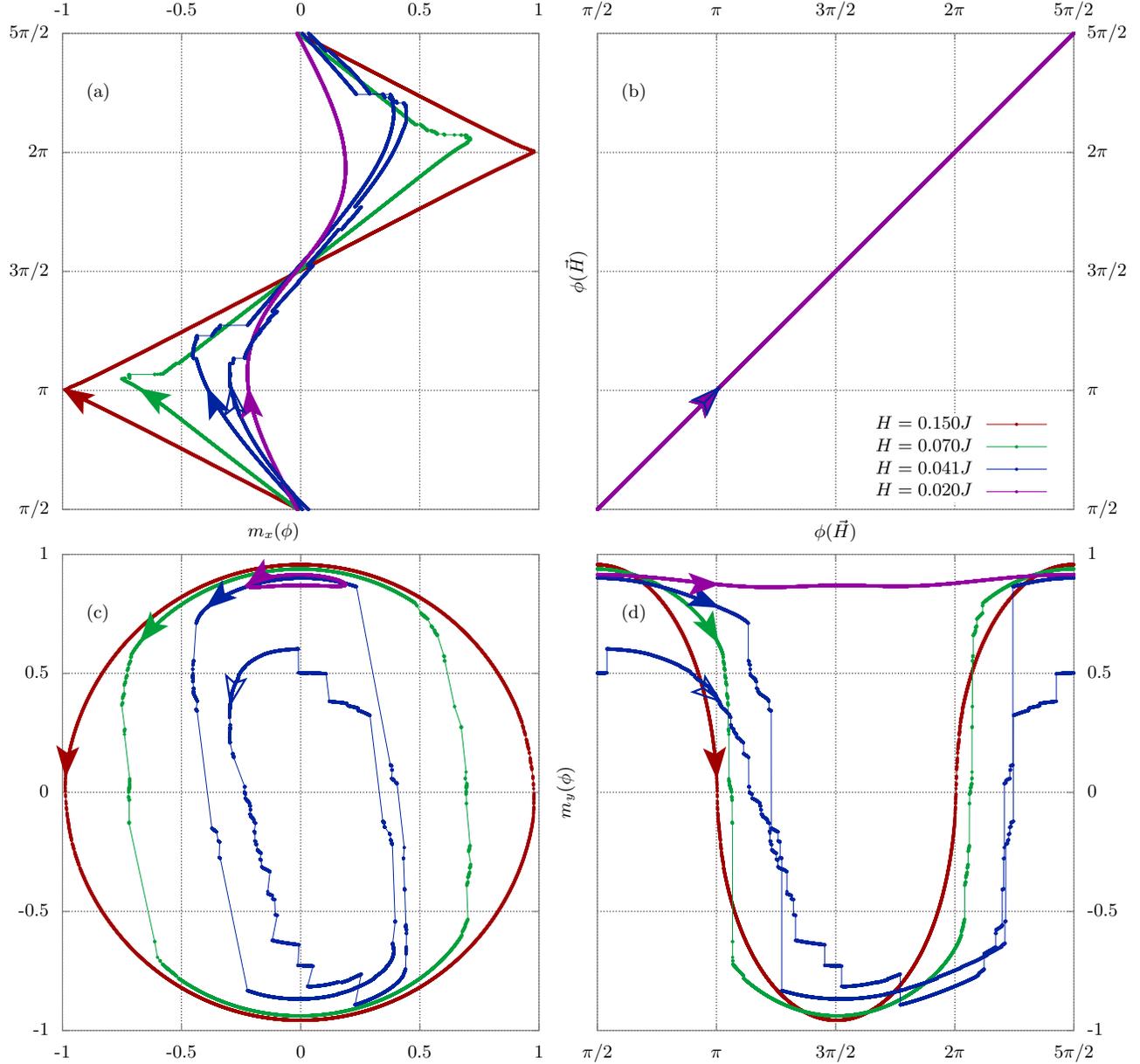}}%
     }
    \caption{ Steady state response to rotating applied field.
A system of size  $ N = 160 \times 160 $ with $R_x=0.5J$ is 
 started from an initial applied field in the $y$-direction.
 The initial spin configuration is aligned with the applied field,
 then relaxed acccording to Eqns.~\ref{eqn:discrete_energy_update}
 as described in the text, after which the applied field 
 is rotated counterclockwise as denoted in panel (b).  
 Panels (a), (c), and (d) show the response once steady state is reached
 under the driving field.
Panel (a) shows the response of the magnetization in the $x$ 
direction, while panel (d) shows the response of the magnetization in the $y$ direction. 
Panel (c) is a parametric plot of $m_y$ {\em vs} $m_x$.
In all panels, the arrows denote the state of the system 
when the driving field is at an angle $\phi = \pi$,
{\em i.e.} aligned along the $x$ direction.
For driving field strength $H = 0.041 J$, the response of the 
system has double the period of the driving field.
The open arrow on this trace denotes the state of the system 
at driving field angle $\phi = \pi$ 
during every other cycle of the driving field.
    }
  \label{fig:4panel_Mx_My_for_h_160}
  \end{center}
\end{figure*}

\section{Results}
\subsection{Behavior of the Limit Cycles}

Fig.~\ref{fig:4panel_Mx_My_for_h_160} shows the rich behavior
of the limit cycles in rotating driving field, as a function of the magnitude of the driving field $H$
at intermediate disorder strength $R_x = 0.5 J$. 
Panel (b) shows the sense of the driving field, which is held at 
constant magnitude, but rotated counterclockwise, {\em i.e.} $\phi$ increases in time
as $\phi = \omega t$ in the $\omega \rightarrow 0$ limit, starting from $\phi = \pi/2$.    
Fig.~\ref{fig:4panel_Mx_My_for_h_160} (a) shows a plot of $m_x$ {\em vs.} the angle $\phi$ of the applied field.
Panel (d) shows a plot of $m_y$ {\em vs.} the angle $\phi$ of the applied field.
Panel (c) shows the combined parametric plot of magnetization $m_x$ in the $x$ direction,
plotted against the magnetization $m_y$ in the $y$ direction.  
The sense of the parametric plot in panel (c) is counterclockwise.  
In each case, the system is started from a locally stable configuration in applied field $\vec{H}||y$ at zero temperature, 
which has been relaxed from an initially saturated state aligned with the initial applied field.   The transient response
before the limit cycle is not shown in this figure.  We discuss the transient response in Sec.~\ref{sec:transient}.

For moderate disorder strength $R_x = 0.5 J$, we find that at small amplitudes
of the driving field, the spontaneous magnetization in the $y$ direction remains robust.
This is evident in the small hysteresis loops we find for $H = 0.02 J$
as shown by the purple trace in the parametric plot Fig.~\ref{fig:4panel_Mx_My_for_h_160}(c).  
This indicates that the system continues to display spontaneous symmetry breaking
in the $y$ direction, retaining its Ising ferromagnetic character in the presence of weak rotating driving field.

As the magnitude of the applied field is increased, there is a change in behavior
from ferromagnetic to paramagnetic response.  
This is evident in the large, almost circular hysteresis loop we find for 
larger 
$H = 0.15 J$, 
as shown by the red trace in the parametric plot Fig.~\ref{fig:4panel_Mx_My_for_h_160}(c).  
This change is consistent with either a crossover in behavior or
a non-equilibrium phase transition at a critical magnitude of the driving field.  
Note that the rotating hysteresis loops at intermediate driving field strengths 
$H = 0.041 J$ and $H = 0.07 J$ 
have rich structure:
Numerous avalanches are evident in these traces.
As we will see in Sec.~\ref{ssxn:Avalanches_near_the_transition}, the
avalanche structure provides further insight into the question of
whether the change from ferromagnetic to paramagnetic response
is a crossover or a phase transition.  
Perhaps the most intriguing feature of the intermediate driving field regime
is that in the blue trace ($H = 0.041 J$), the
limit cycle has {\em double the period of the driving field}.
We find that limit cycles often become multiperiodic at intermediate field strength, 
for large enough system size.  
We explore this region of the phase diagram in more depth in Sec.~\ref{ssxn:period_increase}.

\begin{figure*}[htbp]
  \begin{center}
    \subfloat[ $T\rightarrow 0,\ R_x=0.5 J$ ]{%
    \label{sfig:RFXY_unax_ZT_Rot_stat_max_mean}%
    \resizebox{.5\textwidth}{!}{%
        \input{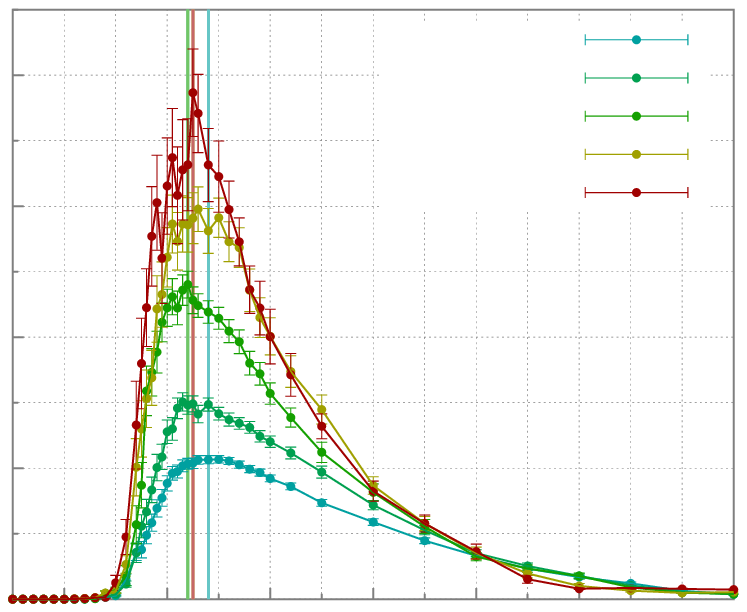}}%
    }
        \subfloat[ $T\rightarrow 0,\ R_x=0.5 J$ ]{%
    \label{sfig:RFXY_unax_ZT_Rot_stat_sq_mean}%
    \resizebox{.5\textwidth}{!}{%
        \input{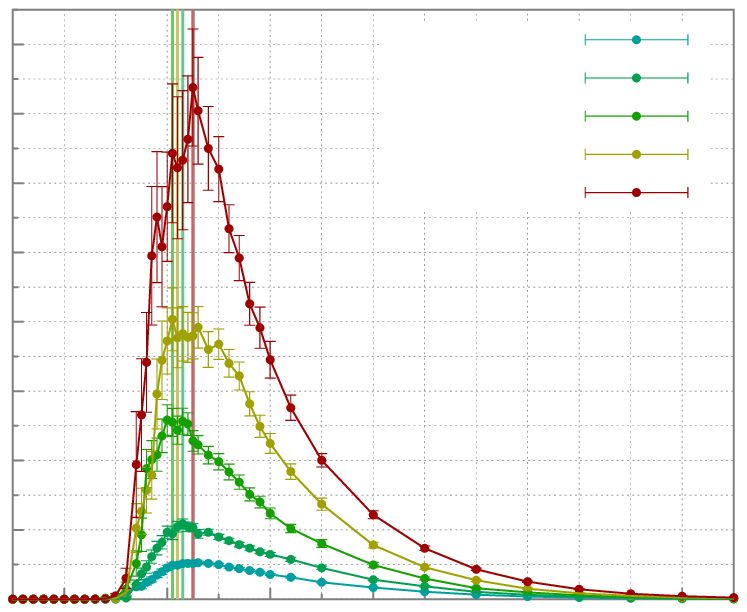}}%
    }
    \caption{ Avalanche statistics.
    The response of the magnetization to rotating driving field often proceeds via avalanches, in which there is a discontinous jump
    in the magnetization $\delta M$ in response to a small change $\delta \phi$ of the driving field angle.  
    In panel (a), we plot the size of the largest avalanche $|\delta \vec{M}|_{\rm max}$ 
    per limit cycle at each rotating field strength, disorder-averaged, for a range of system sizes.
Panel (b) shows the disorder-average of the second moment $\delta \vec{M}$ of the avalanche size distribution.
The brackets $\left< ~~ \right>$ denote an average over the limit cycle,
and the overbar denotes a disorder average.  
By both of these measures, the size of the avalanches grows with system size implying divergent fluctuations 
at a critical field strength in the thermodynamic limit.
The vertical bars in both panels mark the peak value from a running 3-point average.  
Within the resolution of the plot in panel (a), these values are concident for sizes $N = 80 \times 80$ and $N = 100 \times 100$,
and for sizes $N = 128 \times 128$ and $N = 160 \times 160$.  
In panel (b), the peak values are coincident for sizes $N = 64 \times 64$ and $N = 160 \times 160$.
    }
  \label{fig:RFXY_unax_ZT_Rot_stat}
  \end{center}
\end{figure*}

\begin{figure*}[htbp]
  \begin{center}
    \subfloat[ $R_x=0.5 J;~H = 0.058 J;~ N = 64 \times 64. $\label{sfig:tracking_Mx_My_for_h_64}]{%
      \resizebox{.45\textwidth}{!}{%
        \input{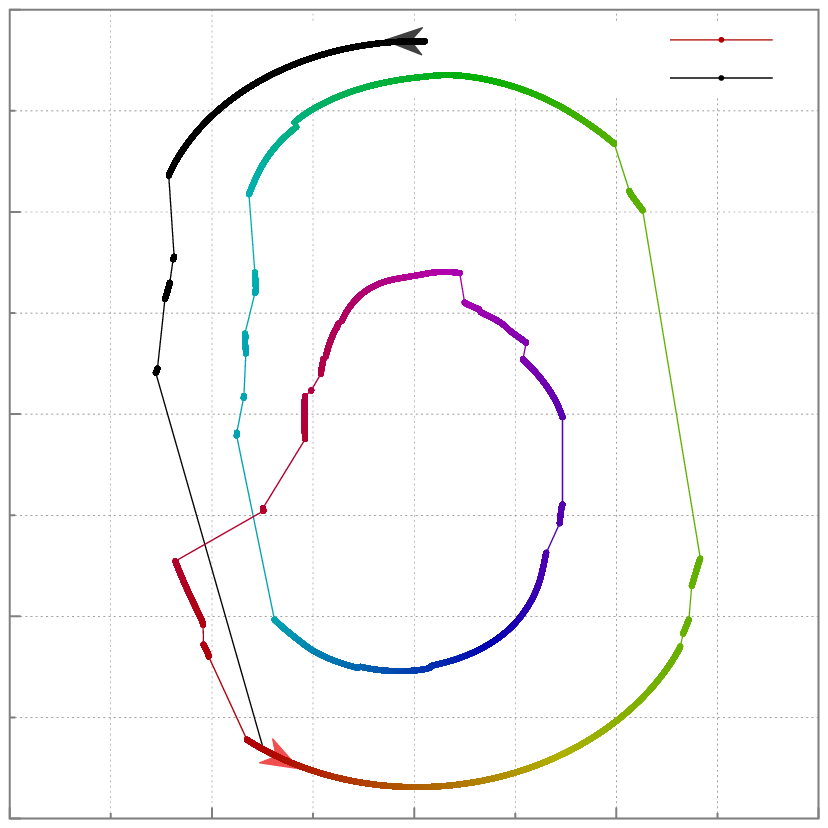}}%
    }%
    \subfloat[ $R_x=0.5J;~ H = 0.048 J ;~ N = 100 \times 100. $\label{sfig:tracking_Mx_My_for_h_100}]{%
      \resizebox{.45\textwidth}{!}{%
        \input{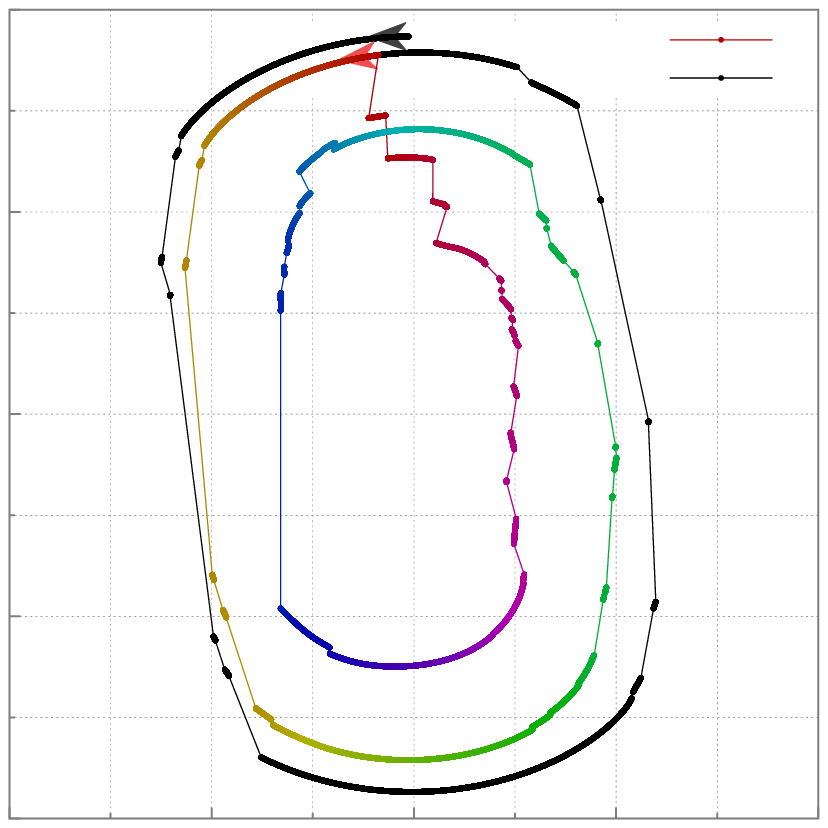}}
    }%
    \\
    \subfloat[ $R_x=0.5 J;~ H = 0.046 J ;~ N = 160 \times 160. $\label{sfig:tracking_Mx_My_for_h_160}]{%
      \resizebox{.45\textwidth}{!}{%
        \input{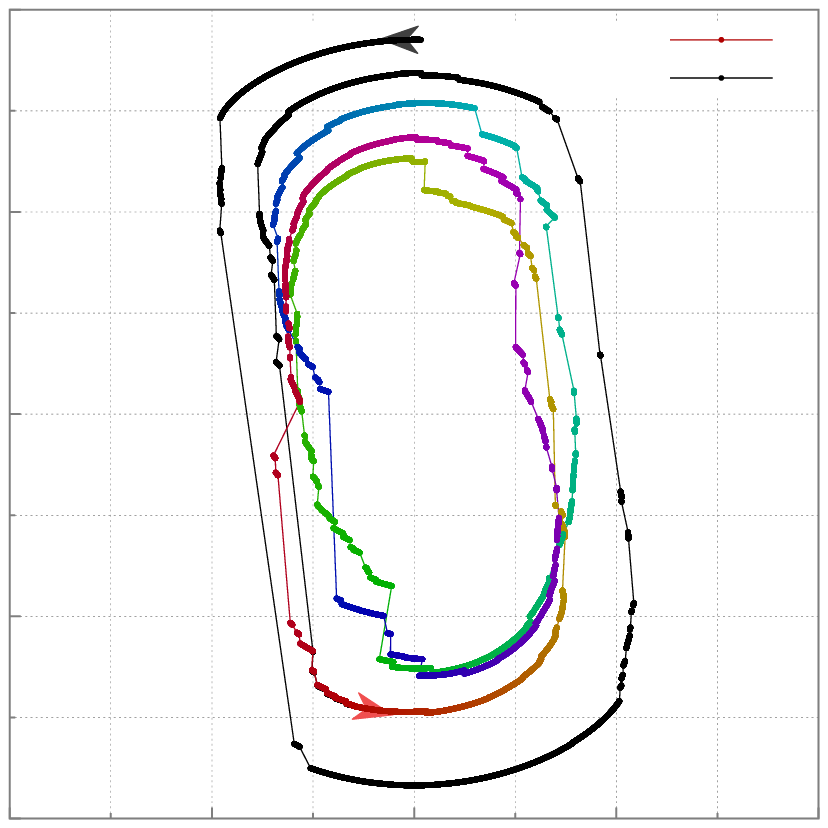}}%
    }
    \subfloat[ $\ R_x=0.5 J,\ R_y=0$ ]{%
    \label{sfig:phi_transient_dis_avg}%
    \resizebox{.45\textwidth}{!}{%
        \input{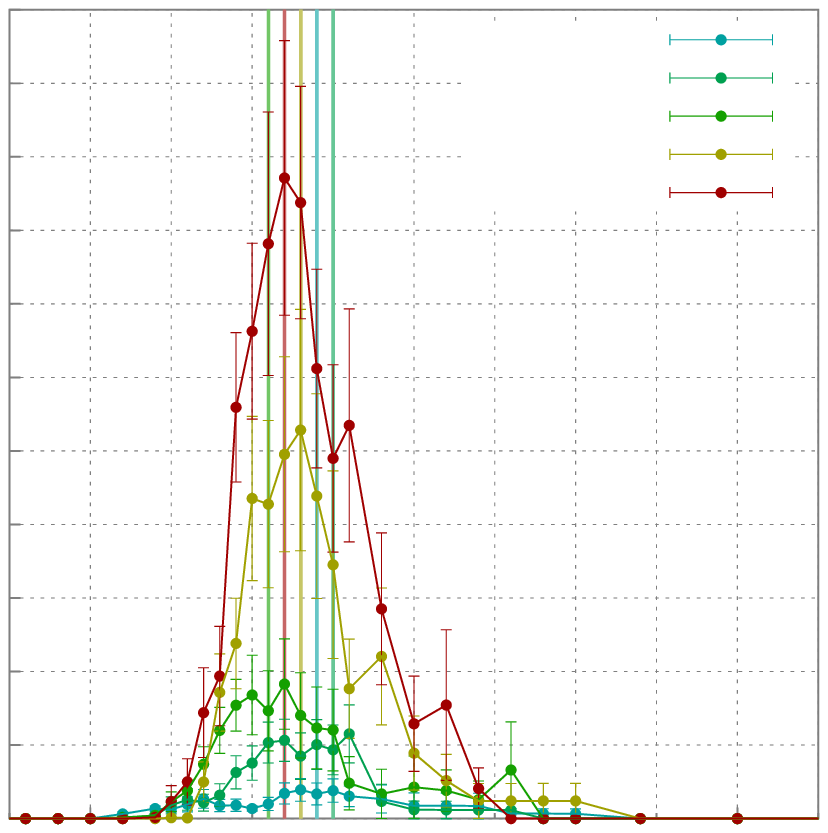}}%
    }%
    \caption{Transient response and multiperiod limit cycles near the transition.
    Panels (a-c) show the initial transient response (black curves), followed by
    multiperiodic limit cycles (rainbow curves). 
    (a) Transient response and multiperiodic limit cycle for one disorder configuration at $N=64^2$.
Here, the transient response lasts roughly half a cycle before a period-2 limit cycle is established.    
    (b) Transient response and multiperiodic limit cycle for one disorder configuration at a larger system size $N=100^2$. 
    Here, the transient response lasts roughly one cycle before a period-2 limit cycle appears.
    (c) Transient response and multiperiodic limit cycle for one disorder configuration at an even larger system size $N=160^2$.
    Here, the transient response lasts almost 1.5 cycles before a period-3 limit cycle is established. 
    (d) The disorder-averaged duration of the transient response, as a function of $H$.  
    The mean of the transient distribution function for each system size is marked by a vertical line of the corresponding color.
    }
  \label{fig:transient}
  \end{center}
\end{figure*}

\subsection{Avalanches Near the Transition}
\label{ssxn:Avalanches_near_the_transition}

In this section we focus on the characteristics of the avalanches that occur near the transition from Ising ferromagnetic to paramagnetic response.
We find a rich avalanche structure at intermediate field strengths, as can be seen in 
the blue and green traces in Fig.~\ref{fig:4panel_Mx_My_for_h_160}
($H = 0.041J$ and $H = 0.07 J$, respectively).
Notice that while the avalanches are apparent in both $m_x$ and in $m_y$, they are
most prominent in $m_y$, which serves as the order parameter in this system.
When magnetization is cast as an extensive quantity, 
$\vec{M}  = \sum_{i=1}^{N} \vec{S}_i = N \vec{m}$,
then in the thermodynamic limit, avalanches $\delta \vec{M}$ of diverging size accompany a second order phase transition.

Fig.~\ref{sfig:RFXY_unax_ZT_Rot_stat_max_mean} plots the size of the largest avalanche $|\delta \vec{M}|_{\rm max}$ at each rotating field strength,
for a range of system sizes $N = L \times L$.  Results are averaged over several disorder configurations of the random field
at disorder strength $R_x = 0.5J$,
ranging from $75$ disorder configurations for system size $N = 64^2$, to $30$ disorder configurations for system size $N = 160^2$.  
(See Appendix~\ref{sec:disorder_averages}.) 
Notice that fluctuations as measured by the largest avalanche diverge with increasing system size at a critical driving field strength, $H_c(R_x = 0.5J)$.
We estimate the value of $H_c$ at $R_x = 0.5J$ as follows:  
For each system size, the peak value based on a 3-point average is indicated by the vertical bar.   
The corresponding peak value of the applied field strength, averaged over all system sizes, 
is $H_c=(0.0452 \pm 0.0015)J$.

In Fig.~\ref{sfig:RFXY_unax_ZT_Rot_stat_sq_mean},
we plot the second moment of all avalanches 
in each limit cycle, $\left< ( \delta \vec{M})^2 \right>$
at each rotating driving field strength, for a range of system sizes.  
Results are disorder averaged, using the same number of disorder configurations
as in Fig.~\ref{sfig:RFXY_unax_ZT_Rot_stat_max_mean}.
Notice that this alternate measure of fluctuations based on the second moment of 
the avalanche size distribution is also consistent with the system undergoing
a second order, nonequilibrium phase transition at a
critical driving field strength, $H_c$.
In this case, we find that $H_c(R_x = 0.5J)=(0.0432 \pm 0.0016)J$,
in agreement with the value of the critical field strength we find  from Fig.~\ref{sfig:RFXY_unax_ZT_Rot_stat_max_mean}.


\begin{figure*}
\centering
\includegraphics[width=\textwidth]{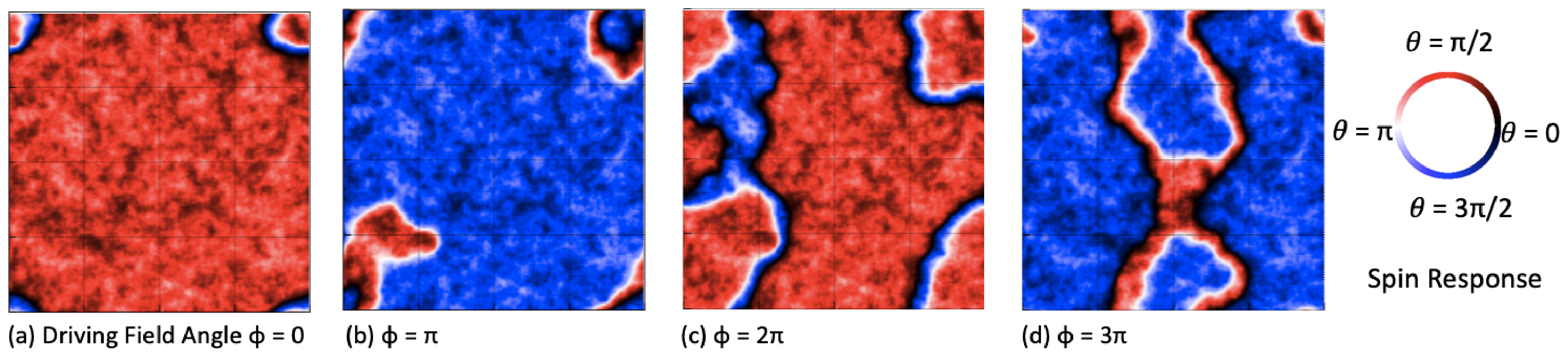}
\caption{
Example of spin configurations during a period-2 limit cycle.  
Spin configurations (a) and (c) are for the same angle $\phi$ of the driving field,
but the spin configuration is different the second time through the driving cycle.
Likewise, spin configurations (b) and (d) are for the same angle $\phi$ 
of the driving field, but the spin configuration is different the second time through the driving cycle.
For this particular disorder configuration and system size, the spin configurations repeat every 2 periods of the driving cycle.
Here, the driving field strength is $H = 0.04J$, and the system size is 160$\times$160.  
See Supplementary Information for further examples, and links to videos.
} \label{fig:phase_plots}
\end{figure*}

\begin{figure*}[htbp]
  \begin{center}
    \subfloat{%
    \label{sfig:multi_period_stats}%
    \resizebox{.5\textwidth}{!}{%
        \input{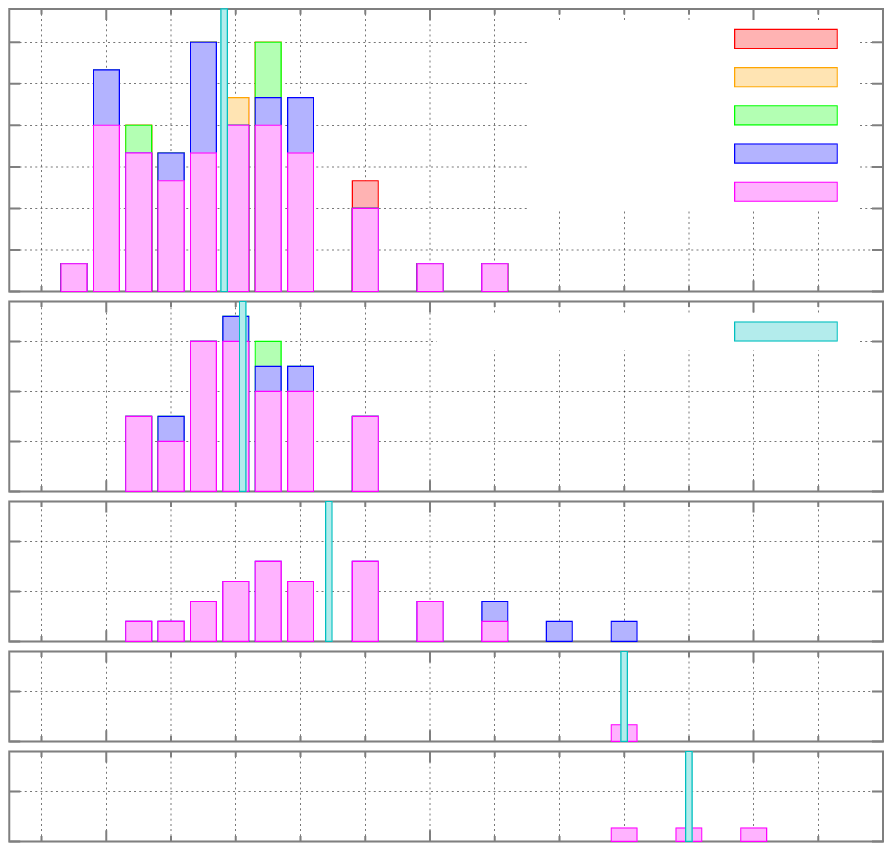}}%
    }%
    \subfloat{%
    \label{sfig:multi_period_avg}%
    \resizebox{.5\textwidth}{!}{%
        \input{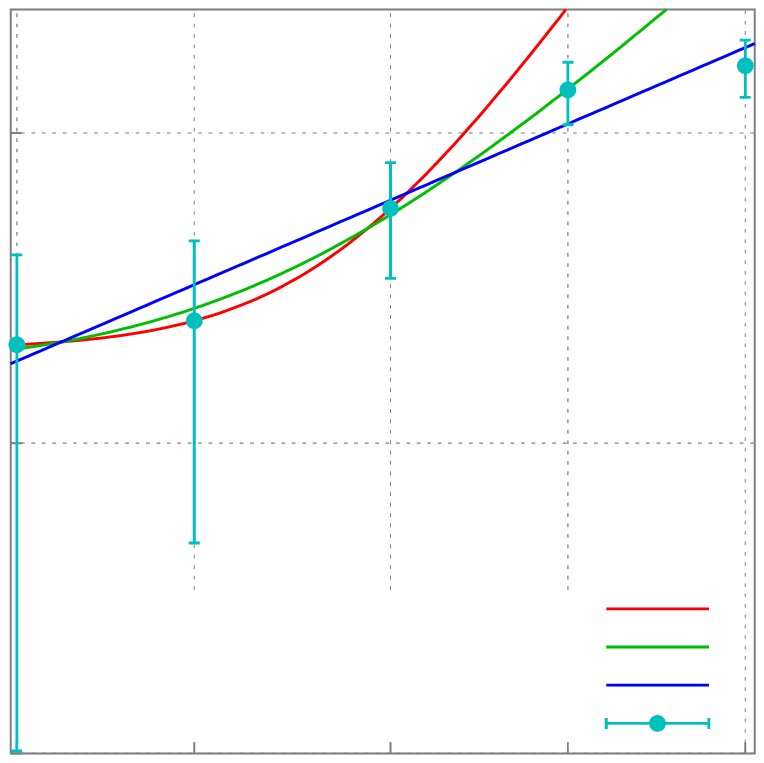}}%
    }
    \caption{Multiperiod limit cycles.  
  Panels (a-e)   show what fraction of limit cycles  exhibit multiperiodicity as a function of driving field strength $H$ at $R_x = 0.5J$.
The smallest system size we simulated, $N = 64 \times 64$, is shown on the bottom  left in panel (e).  
System size increases from bottom to top in the left panels, up to system size $N = 160\times 160$.  
In the bar graphs, period-2 limit cycles are shaded pink; period-3 limit cycles
are  purple; period-4 limit cycles are green;
the period-5 limit cycle is orange; and the period-7 limit cycle is red.
We did not observe any period-6 limit cycles.  
In each bar graph, the vertical blue line is the mean of the distribution function, $\left< H_{lc} \right>$ in units of $J$. 
    (f) From the results of panels (a-e), we plot $ \langle H_{lc} \rangle _{N} $ {\em vs.} the inverse of system size $N$ on a log-log scale. 
    A power law fit of $\langle H_{lc} \rangle_N$ for the three largest system sizes is given by the red curve; the fit for the four largest system sizes is given by the green curve; and the fit for all calculated system sizes is given by the dark blue curve. 
    The $y$-intercept is consistent among all of these fits, yielding an average value of 
    $ \langle H_{lc} \rangle_{N \to \infty} = (0.0434\pm 0.0020)J$ .
    }
  \label{fig:limitcycle}
  \end{center}
\end{figure*}

\subsection{Transient Response}
\label{sec:transient}

  Fig.~\ref{fig:transient}(a-c) shows how the magnetization responds to a rotating driving field
  in the vicinity of the phase transition.  
  There is a transient response before the system settles into a limit cycle.
  A limit cycle is the steadily repeating response in the magnetization due to a rotating driving field. 
  While we find that most limit cycles have the same period as the driving field, 
  we find that near the transition regime,  limit cycles often have a longer period.
We first discuss the behavior of the transient response, before turning our attention to
the behavior of the multiperiodic limit cycles in Sec.~\ref{ssxn:period_increase}

The transient response in panels (a-c) of  Fig.~\ref{fig:transient} is marked in black.
In  Fig.~\ref{fig:transient}(d), we plot the duration of the transient response,
as a function of $H$, for various system sizes.  The results shown have been averaged
over several disorder configurations.  (See Appendix~\ref{sec:disorder_averages} for details.)
At high and low strength of the driving field,
the transient response becomes so negligible
as to be smaller than the symbol size on this graph.
However, at intermediate driving field strength, the transient response 
grows with increasing system size.
The fact that the transient response grows with increasing system size is further corroboration
that the system is undergoing a second order phase transition.     
In Fig.~\ref{fig:transient}, 
the mean of each transient distribution function is denoted by a vertical line,
color coded to the system size.  
The average of the mean value of $H$ from these  vertical lines is 
 $ \langle H_{tr} \rangle_{N} = (0.0430 \pm 0.0014)J$, consistent with our previous estimates of $H_c(R_x=0.5J)$.

\begin{figure*}[htbp]
  \begin{center}
      \subfloat[ Maximum period of limit cycles. \label{sfig:Non_repeatability_max_loop}]{%
    \resizebox{.5\textwidth}{!}{%
        \input{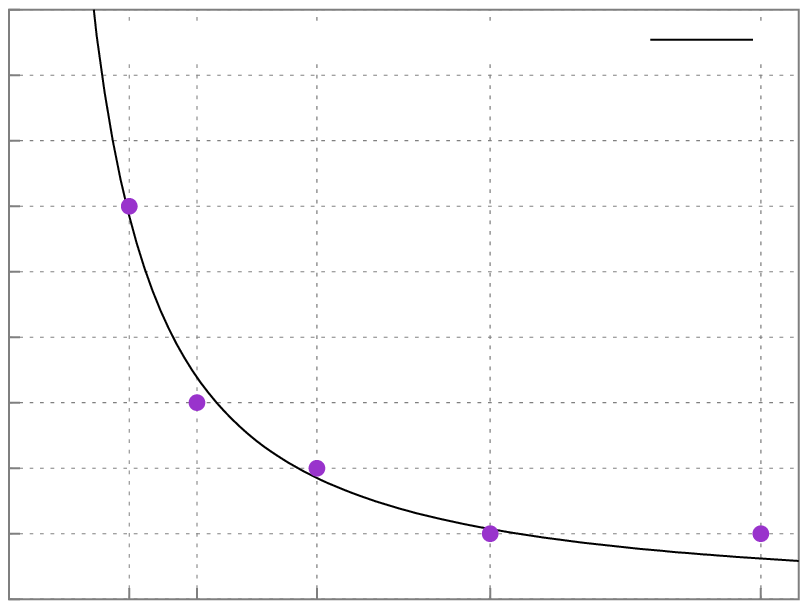}}%
    }
    \subfloat[ Maximum likelihood of multiperiod limit cycles. \label{sfig:Non_repeatability_max_prob}]{%
    \resizebox{.5\textwidth}{!}{%
        \input{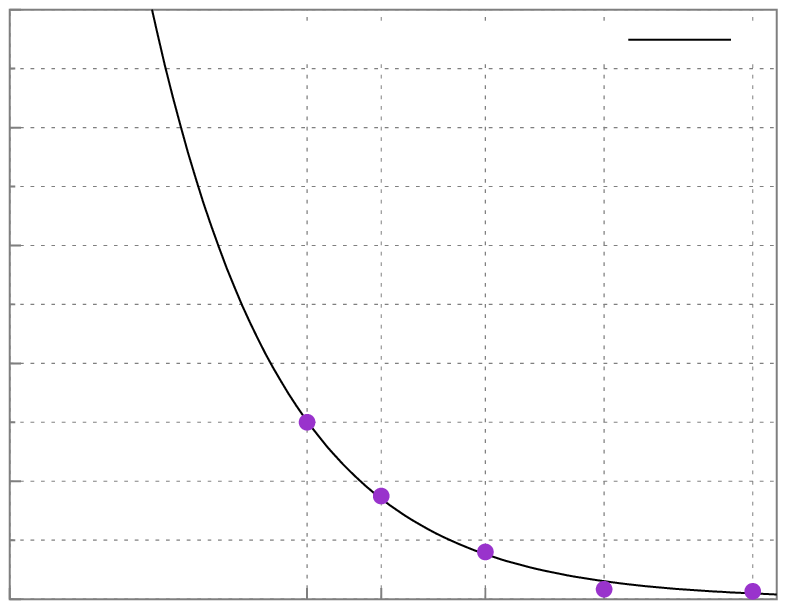}}%
    }%
    \caption{
 Trends of the multiperiodic behavior of the limit cycles with increasing system size.
In panel (a) we plot the
 maximum period of the limit cycles observed in
Fig~\ref{fig:limitcycle}(a-e), as a function of 1/N (purple circles).
The black line is a fit to the simulation results. 
 The trend is toward divergence of the period of limit cycles in the thermodynamic limit.
 In panel (b), we plot 
 the maximum likelihood of multiperiodic
 limit cycles, obtained from the peak heights of 
 the left-hand panels in Fig~\ref{fig:limitcycle} (purple circles).  
 The black line is a fit to the simulation results.  
The trend is toward saturation of the likelihood of multiperiod behavior in the thermodynamic limit.
    }
  \label{fig:Non_repeatability}
  \end{center}
\end{figure*}

\subsection{Period Increase Near the Transition}
\label{ssxn:period_increase}

We now turn our attention to the  behavior of the limit cycles at intermediate driving field strength.
One of the most fascinating features of the limit cycles in this regime
is that some of them have a longer period than that of the driving  field.
Fig.~\ref{fig:transient}  shows some representative cases of this behavior.
Fig.~\ref{fig:phase_plots} visualizes how the spin configurations respond 
to the driving field during one of the period-2 limit cycles.  Domain walls have dramatically
different configurations during the second cycle as opposed to the first cycle of the driving field,
suggesting a prominent role for domain wall pinning and domain wall creep.   
More examples of such behavior, including links to videos of spin configurations
during multiperiod limit cycles, can be found in the Supplementary Information.

In order to explore this behavior quantitatively, we studied several
disorder configurations near the transition, as a function of system size.
Fig.~\ref{fig:limitcycle} shows a histogram of the likelihood of multiperiod limit cycles.
For a given magnitude of the driving field $H$ and a given system size $N$,
we plot the number of disorder configurations whose limit cycle has a period
greater than that of the driving field, divided by the number of all disorder
configurations studied at that $H$ and $N$.  
Starting from the bottom panel on the lefthand side of Fig.~\ref{fig:limitcycle}, panel (e),
the system size increases as one moves to the next panel up the page,
up to panel (a) which shows the largest system we studied, $N = 160 \times 160$.  
Different color bars indicate the period of the multiperiod behavior:
pink indicates period doubling; blue shows period tripling;
period-4 limit cycles are denoted in green; yellow is for period-5, and orange is for period-7.
We did not observe any period-6 limit cycles, although presumably
these would appear at certain disorder configurations as well.

The vertical blue bars mark the mean of the distributions in
Fig.~\ref{fig:limitcycle}(a-e),
$ \langle H_{lc} \rangle_{N}$.
In Fig.~\ref{fig:limitcycle}(f), we plot $ \langle H_{lc} \rangle_{N}$ 
vs $1/N$ on a log-log plot, in order to determine the limiting 
value $ \langle H_{lc} \rangle_{N \to \infty}$.
Fits of the finite size scaling in Fig.~\ref{fig:limitcycle}(f)
for all system sizes, the four largest system sizes,
and the three largest system sizes yield a consistent
value for $ \langle H_{lc} \rangle_{N \to \infty}$ within error bars.
The average of these three methods yields $ \langle H_{lc} \rangle_{N \to \infty} = (0.0434 \pm 0.0020)J$.

\begin{figure}[htbp]
  \begin{center}
  \subfloat{%
    \resizebox{.5\textwidth}{!}{%
      \input{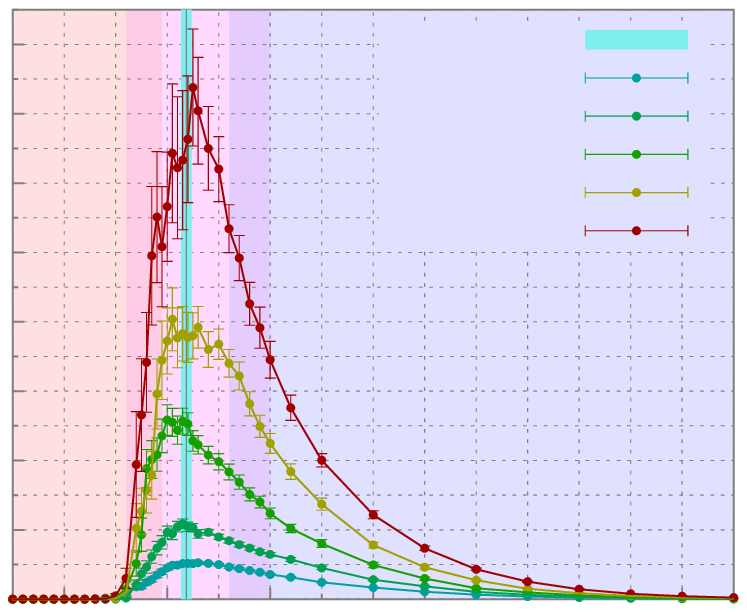}}%
  }
  \caption{
    This figure shows the phase diagram as a function of the strength of the rotating field in a uniaxial Random field. The region where the number of multiperiodic loops and the maximum periodicity increases with system size is labeled as the expected region of non-repeatablity for infinitely large systems. This region coincides with the region where the largest avalanche occurs is this system where $\langle H_c \rangle = (0.0437 \pm 0.0009)J$, which is marked by the vertical blue line.
  }
  \label{fig:phase_diagram_uaxRF_rotating_field}
  \end{center}
\end{figure}

\subsection{Approach to Non-Repeatability}
\label{ssxn:non_repeatability}

We find that at small system size, multiperiod behavior is rare. 
However, as the system size is increased, and the disorder configurations
can become correspondingly more rich, the likelihood of 
multiperiod behavior increases.
In Fig.~\ref{fig:Non_repeatability}(a), we
plot the maximum observed period of a limit cycle, {\em vs.}
$1/N$.    The maximum period increases with increasing system size,
in a manner consistent with diverging period in the thermodynamic limit.

Notice also that the distribution in Fig.~\ref{fig:limitcycle}(a-e) grows in height 
with increasing system size. 
For $N = 160 \times 160$, we find that $20-30 \%$ of disorder
configurations in the range $H = (0.04 - 0.046)J$ display multiperiodic behavior. 
To quantify these effects, we plot the maximum height of the distributions in
Fig.~\ref{fig:limitcycle}(a-e) in 
Fig.~\ref{fig:Non_repeatability}(b).  This measure also shows sharp
increase with increasing system size.
The fact that both the likelihood
of multiperiod behavior and the period of limit cycles 
steadily increase with increasing system size
points toward a thermodynamic limit in which 
the period of limit cycles goes to infinity.
If the period of a system diverges in the thermodynamic limit,  
then the system has effectively entered a regime
of non-repeatability.  We discuss further implications of this finding in the next section.

\section{Discussion}
\label{Evidence_of_Non_repeatability}

Using four different methods to quantify the fluctuations in the system
 (see Table~\ref{tab:hc}),
we find evidence for a second order nonequilibrium phase transition
from  spontaneous Ising ferromagnetism at low driving field strength,
to XY paramagnetism at high driving field strength.
The critical field strength at which this transition occurs
is consistent across all methods we employed, yielding
an average value 
 of $H_c = 0.0437 \pm 0.0009$,
as denoted in the phase diagram in Fig.~\ref{fig:phase_diagram_uaxRF_rotating_field}.

\begin{table}[ht]
\caption{Critical field strength.  } 
\centering 
\begin{ruledtabular}
\begin{tabular}{c c } 
Method & Value of $H_c/J$  \\ [0.5ex] 
\hline 
{Largest avalanche of limit cycle} & $0.0452 \pm 0.0015$ \\ 
{Second moment of avalanches in limit cycle}  & $ 0.0432 \pm 0.0016$ \\ 
{Duration of transient response}  & $0.043 \pm 0.0014$ \\ 
{Finite size scaling of multiperiodic behavior} & $0.0434 \pm 0.0020$ \\
\hline
{Overall average of above methods} & $0.0437 \pm 0.0009$  \\
\end{tabular}
\end{ruledtabular}
\label{tab:hc}
\end{table}

We furthermore find that far from being irrelevant, disorder plays a prominent role at the transition.
Because the disordered energy landscape makes the system 
highly susceptible to spatial fluctuations near the transition,
there is both a longer transient response and a 
longer period of  limit cycles near $H_c$.  
Remarkably, both the likelihood of multiperiod behavior
and the period of the limit cycles increases with no sign of saturation
as system size is increased. 
The trend we find is toward a thermodynamic limit in which
limit cycles never repeat.  
A large enough physical system at this critical point should therefore
display a regime of {\em non-repeatability}.
As shown in  Fig.~\ref{fig:phase_diagram_uaxRF_rotating_field},
the regime of non-repeatability in the thermodynamic limit coincides with the nonequilibrium phase transition. 
The dependence of this simple model upon history 
implies that experiments on XY systems in
uniaxial random field are particularly sensitive to disorder.
Conflicting experimental results could arise if 
hysteresis protocols are not closely monitored.

Similar behavior is predicted to occur 
in models of amorphous solids under periodic shear stress.\cite{reichhardt-chaos,reichhardt-dahmen,leishangthem_yielding_2017}.  
In these systems, simulations revealed that under periodic shear,
the response of the system becomes multiperiodic, in a way
that is consistent with chaotic behavior at a critical shear amplitude.  
More work would be needed to determine whether the
multiperiodic cascade observed here is indicative of chaotic
behavior in the thermodynamic limit.
Similar multiperiod cascades signal the onset of chaos 
in nonlinear systems, 
suggesting that the multiperiod cascades 
observed here and in periodically driven models of amorphous solids
are characteristic 
of a larger class of transitions in dynamical systems.

On the other hand, we predict that {\em finite size} physical systems in the vicinity of
the nonequilibrium transition should display the characteristics of a 
classical discrete time crystal,\cite{yao-nayak-2018,wilczek_cdtc_2012} in which the discrete time translation symmetry
imposed by the periodic drive is broken in a way that leads to
rigid subharmonic entrainment. 
We find that the period of the response remains stable against perturbations
in the initial conditions and stable against low temperature fluctuations (see the SI), indicating that the 
spontaneous breaking of the discrete time symmetry is rigid.
Yao {\em et al.} find that the critical endpoint between
a classical discrete time crystal and the disordered phase
of a dissipative, coupled chain of classical nonlinear pendula
terminates in a critical point which is not in the Ising universality class.
Because the nonequilibrium transition we find here is in the Ising universality class,
this indicates that there is more than one classical discrete time crystal universality class.
The results here further underscore the fact that long-range interactions
are not a necessary ingredient to stabilize a time crystal.\cite{yu-short-range}

The work in this paper was done at uniaxial random field strength $R_x = 0.5J$,
with zero random field strength in the $y$-direction.
Further work is needed to obtain the full phase diagram as a function of
random field strengths $R_x$ and $R_y$.


\subsection{Applications to Physical Systems}
\label{sxn:applications}

The uniaxial random field XY model has been applied to many systems, including
layers of Josephson junctions,\cite{lacour-gayet_toulouse}
superfluid in a uniaxially stressed aerogel,\cite{PhysRevB.73.060504} 
ultracold atoms in the presence of speckle radiation,\cite{PhysRevB.74.224448}
uniaxially stressed 2D Wigner crystals,\cite{PhysRevB.23.335,PhysRevB.19.2457,PhysRevLett.41.121,PhysRevB.19.1855} 
and the half-integer quantum Hall effecte\cite{Lilly-1999-PhysRevLett.82.394} 
Uniaxial random field-induced has also been discussed in 
connection with the graphene quantum Hall ferromagnet.~\cite{abanin-lee,girvin-macdonald-book}
We discuss below a few of these systems in which 
there is also a clear way to drive the system with a rotating field.

\subsubsection{Electron nematics}
\label{sxn:nematics}

An electron nematic occurs when the electronic degrees of freedom spontaneously break
the rotational symmetry of the host crystal.
Electron nematics have been observed or proposed in several material systems,
including transition metal oxides like cuprate superconductors, manganites, nickelates,
and cobaltites;  valley symmetry breaking systems like single and bilayer graphene, elemental bismuth,
and AlGaAs 2DEG's, as well as strontium ruthenates and iron pnictides.\cite{kivelson-rmp-2003,kivelson-science-2010}
In mapping any nematic to an XY model, there is a factor of two between the
physical angle of the nematic in the plane, and the natural angles in an XY model.
This is because a nematic is symmetric under $180^o$ rotation, whereas
the XY spins change sign under the same operation.
For an electron nematic arising out of a crystal with discrete C4 rotational symmery, 
the nematic order parameter switches sign when the nematic rotates by $90^o$. 
The uniaxial random fields we discuss in this paper can
arise in these systems if random orienting fields are strong only along the
major crystalline axes.    Note that in this case, the order-by-disorder
transition would induce the electron nematic to orient along a direction
which is {\em diagonal} to the major crystalline axes.  

Several external perturbations can be used as a driving field on an 
electron nematic, including magnetic field, electric field, high currents,
and uniaxial stress.\cite{carlson-dahmen-natcomm-2011}
Note that similar symmetry considerations apply to
the driving field in these systems.
For example, a rotating applied magnetic field 
$\vec{B} = [B_x, B_y] =  B [{\rm cos}(\omega t)), {\rm sin}(\omega t)] $
can be used to
exert the rotating driving field of Eqn.~\ref{eqn:model_XY_uniax_RF} for the case of a nematic, 
with the
caveat that rotating the applied field by $90^o$ 
changes the sign of the driving field:
\begin{equation}
\vec{H} = [H_x,H_y] = H[{\rm cos}(2 \omega t), {\rm sin} (2 \omega t)]~.
\end{equation}

%
%

\subsubsection{Quantum Gases}

Random-field induced order has been proposed to happen in 
coupled Bose-Einstein condensade systems.\cite{PhysRevB.74.224448}
Theoretical and numerical results on two-component Bose gases
predict that by using a Raman field to couple two internal states,
uniaxial random field disorder can be produced.  
The uniaxial nature is achieved by a Raman coupling with constant phase,
while the randomness is achieved through random strength of the Raman field.\cite{niederberger-prl-2008,sanchez-palencia-natphys-2010}
Similarly, a rotating driving field can be applied by a Raman coupling with
uniform strength, but rotating phase.

\subsubsection{Magnetic systems}
While the mapping of a magnetic system with XY symmetry to
Eqn.~\ref{eqn:model_XY_uniax_RF} is clear, 
the realization of a uniaxial random field in these systems is less clear.
It may be possible to design a system in which 
epitaxial strain from a substrate exerts random uniaxial fields
on a 2D XY ferromagnet through magnetoelastic coupling.

%
%

\subsection{Conclusions}
We have shown that the order-by-disorder transition of the 
two-dimensional XY model in the presence of a uniaxial random field
persists up to a critical strength of the rotating driving field.
Near the critical driving field strength, 
the response of the system has a period which is
an integer multiple $n>1$ of the driving field period.
This spontaneous breaking of the discrete time symmetry of the 
driving field indicates that a classical discrete time crystal
emerges at the critical point.    
The trend with increasing system size is toward increasing period $n$,
indicating both that the period of the time crystal can be engineered in small systems,
and also suggesting the onset of what is effectively non-repeatability 
as $n\rightarrow large$ in the thermodynamic limit.
Similar multiperiod cascades signal the onset of chaos 
in nonlinear systems, and signal the onset of irreversibility
in periodically driven models of plastic deformation, 
suggesting that multiperiod cascades are characteristic 
of a larger class of transitions in dynamical systems.



\section{Methods}
\label{sec:methods}


The magnetization $m_y$ in the $y$-direction at intermediate disorder strength $R_x = 0.5J$
remains ordered even in the presence of weak applied transverse field $H_x$
(see Supplementary Information). 
Therefore, to begin the hysteresis studies, we first initialize the system in a $y$-magnetized state, by starting
from the fully saturated $y$ magnetization, with the driving field aligned along $y$,
$\vec{H} || y$, then allow the system to relax\cite{PhysRevE.59.1355} at that applied field.
We take the angle $\phi$ of the applied field to be $\phi = {\rm Arctan}(H_y/H_x)$,
so the initial direction of the applied field is $\phi = \pi/2$.  
After rotating the applied field by an amount $ \delta \phi(\vec{H}) $,
the spin configuration is updated successively so as to minimize the energy,
in the $\omega \rightarrow 0$ limit.  
After a transient response, the response of the system then settles into a limit cycle.  

  Each time the applied field direction is updated, 
  the energy is minimized on each site by aligning the spin on each site with its effective field, $\vec{h}^{eff}_i$. 
  Hence the following update strategy is repeated until the spin configuration converges to the nearest energy minimum:
\begin{eqnarray}
  \vec{h}^{eff}_i(t) &=& J\sum_{j\in\langle i,j \rangle}\vec{S}_j(t)+\vec{h}_i+\vec{H},\nonumber \\
  \vec{S}_i(t+1) &=& \frac{\vec{h}^{eff}_i(t)}{|\vec{h}^{eff}_i(t)|}
  \label{eqn:discrete_energy_update}
\end{eqnarray}
This update mechanism is similar to Eqn. (2) of Ref.~\cite{PhysRevE.83.011121}, 
however the effective on-site field in our case includes only the instantaneous influence of
nearest neighbors, whereas Ref.~\cite{PhysRevE.83.011121} is working in a mean-field limit.
The update algorithm we employ is described in more detail below,
in Sec.~\ref{sec:spin_relax}.

We continue to allow spins to relax under the influence of 
Eqns.~\ref{eqn:discrete_energy_update} until a limit cycle is reached, defined by $\{\vec{S}_i\}(\phi+2\pi n )=\{\vec{S}_i\}(\phi)$.
We use the following parameters in our simulations:  
$\delta m_{cutoff} = 10^{-4}$,
   $\delta \phi_{max}=2\pi\times 10^{-4}$, 
   $\delta \phi_{min} = 2^{-14} \times \delta \phi_{max}$.
Hence the avalanches ($\delta m$) are only well-defined within the precision
of the driving field angle,  
$\delta \phi_{min} = 2\pi \times 6.1 \times 10^{-9}$.  

\subsection{Spin Relaxation Method}
\label{sec:spin_relax}
The rotation of the driving field and subsequent relaxation of the spin configuration is performed as follows.
Starting from an initial spin state $\{\vec{S}_i\}(\phi)$ for a given applied field direction 
$\phi = {\rm Arctan}(H_y/H_x)$
and with $\delta \phi$ initially set to $\delta \phi=\delta \phi_{max}$:
\begin{enumerate}
\item{\label{step:updatephi}}  Update $\phi \rightarrow \phi + \delta \phi$.
\item{\label{step:relax}}  Use Eqns.~\ref{eqn:discrete_energy_update} to relax the spin configuration.
  \item {\label{step:binary_division_3}} If $\delta m > \delta m_{cutoff}$, then:
	  \begin{enumerate}
	  	\item {\label{step:binary_division_3b}} If $\delta \phi = \delta \phi_{min}$, accept the new spin configuration and the new $\phi$, and proceed to Step~\ref{step:updatephi}
    		\item {\label{step:binary_division_3a}} Else reject the changes.   Set $\delta \phi \rightarrow \delta\phi / 2$ and proceed to Step~\ref{step:updatephi}    
          \end{enumerate}
  \item {\label{step:binary_division_4}} Else accept the new spin configuration and the new $\phi$, and:
  \begin{enumerate}
    \item {\label{step:binary_division_4a}} If $\delta \phi = \delta \phi_{max}$ or $\delta m  \ge \frac{\delta m_{cutoff}}{2}$, proceed to Step~\ref{step:updatephi}.
    \item {\label{step:binary_division_4b}} Else, set $\delta\phi\to 2\times\delta\phi$ and proceed to Step~\ref{step:updatephi}.
  \end{enumerate}
\end{enumerate}

\subsection{Disorder Averages}
\label{sec:disorder_averages}

Table~\ref{tab:disorder_average} reports the number of disorder configurations
used in  Figs.~\ref{fig:RFXY_unax_ZT_Rot_stat}, ~\ref{fig:transient}, and~\ref{fig:limitcycle}.

\begin{table}[ht]
\caption{Number of disorder  configurations used in Fig.~\ref{fig:RFXY_unax_ZT_Rot_stat} (a) and (b), Fig.~\ref{fig:transient}(d), and Fig.~\ref{fig:limitcycle} (a-e).}
\centering 
\begin{ruledtabular}

\begin{tabular}{c c } 
{Size ($N = L \times L$)} & Configurations  \\ [0.5ex] 
\hline 

{$64 \times 64$} & {$75$} \\

$80 \times 80 $ & $60$\\ 
$100 \times 100$ & $50$\\
$128 \times 128$ & $40$\\
$160 \times 160$ & $30$\\
\end{tabular}

\end{ruledtabular}
\label{tab:disorder_average}
\end{table}


\section{acknowledgements}
We acknowledge helpful conversations with D.~Campbell, S.~Choudhury, G.~A.~Csathy, E.~Fradkin, T.~C.~Li, and 
C.~Nayak.
E.W.C. and S.B. acknowledge support from NSF Grant No. DMR-1508236, Department of Education Grant No. P116F140459, and XSEDE Grant Nos. TG-DMR-180098 and DMR-190014.
S.B. acknowledges support from a Bilsland Dissertation Fellowship. 
K.A.D. acknowledges support from 
NSF Grant No. CBET-1336634.  
This research was supported in part through computational resources provided by Information Technology at Purdue, West Lafayette, Indiana.

\bibliographystyle{apsrev4-1}
\bibliography{RFXY} 

\end{document}


\title{Supplemental Material for ``Emergent Time Crystal with Tuneable Period in the Uniaxial Random Field XY Magnet''}

\author{S. Basak}
\affiliation{Department of Physics and Astronomy, Purdue University, West Lafayette, IN 47907, USA}
\affiliation{Purdue Quantum Science and Engineering Institute, West Lafayette, IN 47907, USA}
\author{K.~A. Dahmen}
\affiliation{Department of Physics, University of Illinois at Urbana-Champaign, Urbana, IL 61801, USA}
\author{E.~W. Carlson}
\email{ewcarlson@purdue.edu}
\affiliation{Department of Physics and Astronomy, Purdue University, West Lafayette, IN 47907, USA}
\affiliation{Purdue Quantum Science and Engineering Institute, West Lafayette, IN 47907, USA}

\date{\today}

\maketitle

\subsection{Spin configurations under driving rotating field}
\textcolor{black}{In our zero temperature simulations,}
energy is minimized for each site based on the local field and the configuration of the nearest neighbor interactions.
We use two types of driving protocol: one is changing the driving field angle ($\phi$) at a constant rate; the other one is a variable rate where the rate is slowed down if the change in response magnetization is large and sped up if the response is small.
Both these protocol gives us the same periodicity of the limit cycle.
For example, if the constant rate is too large it can merge two avalances into one but the overall magnetization remains the same.


We also observe that the system falls into the same limit cycle however we initialize the spins.
Due to the emergent Ising symmetry in the system and the above observation the 
limit cycles will be the same irrespective of the sense of rotating driving field. 
This is because the spin configurations can be mapped by a symmetry transformation 
from the response limit cycle of a clockwise rotating field to the response 
limit cycle of counter-clockwise rotating field. 
\textcolor{black}
{Only the transient response depends on the initial spin configuration.}

\begin{figure*}[htbp]
  \begin{center}
  \vspace{-5mm}
  \subfloat[$\phi=0$]{%
    \includegraphics[trim={0.5cm 0.8cm 0.5cm 1.2cm},clip,width=0.2\textwidth]{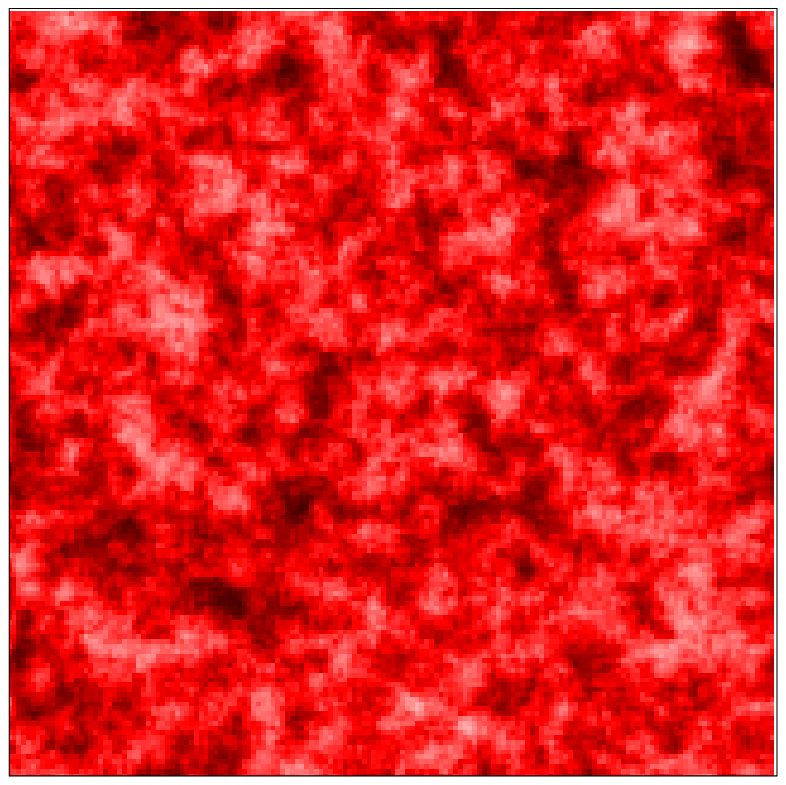}%
  }
  \subfloat[$\phi=0.2\pi$]{%
    \includegraphics[trim={0.5cm 0.8cm 0.5cm 1.2cm},clip,width=0.2\textwidth]{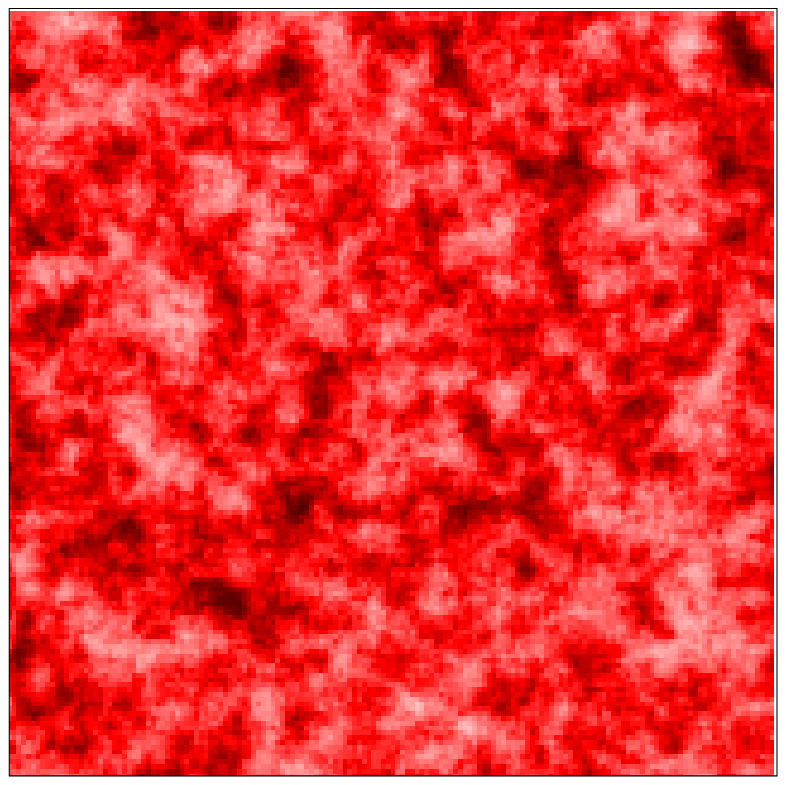}%
  }
  \subfloat[$\phi=0.4\pi$]{%
    \includegraphics[trim={0.5cm 0.8cm 0.5cm 1.2cm},clip,width=0.2\textwidth]{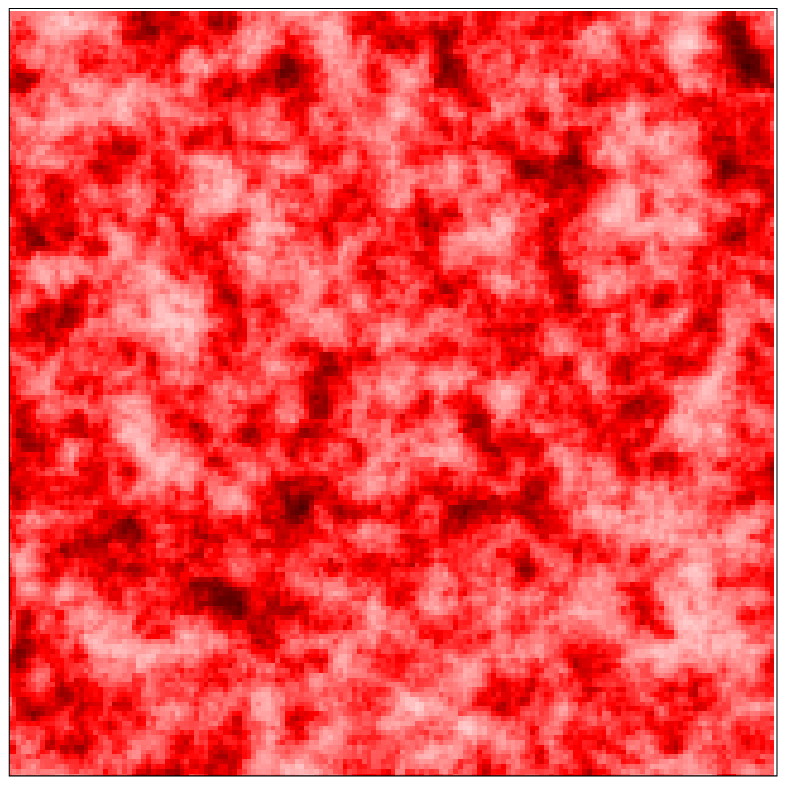}%
  }
  \subfloat[$\phi=0.6\pi$]{%
    \includegraphics[trim={0.5cm 0.8cm 0.5cm 1.2cm},clip,width=0.2\textwidth]{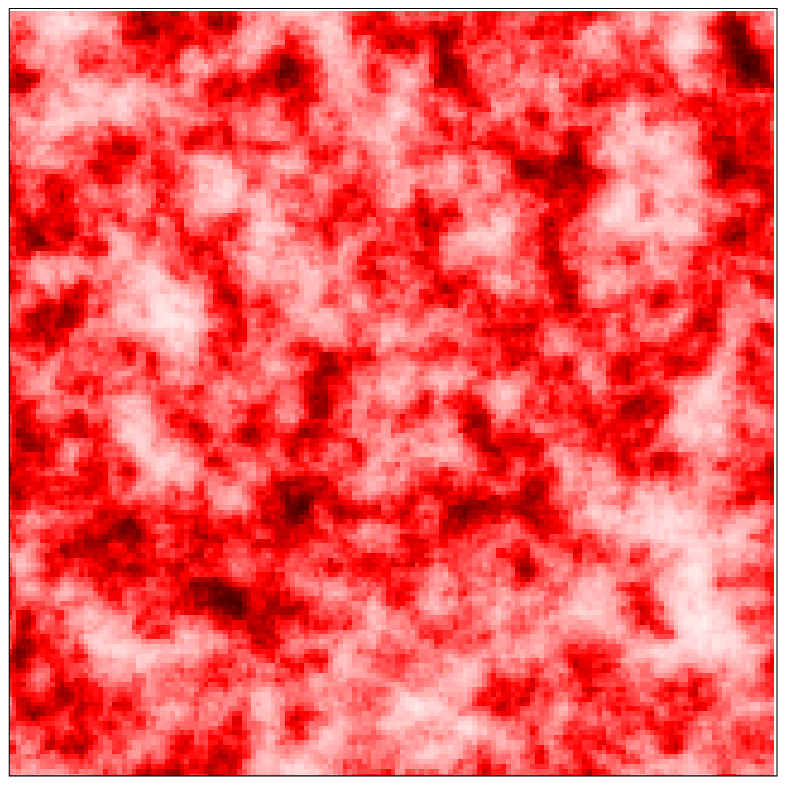}%
  }
  \subfloat[$\phi=0.8\pi$]{%
    \includegraphics[trim={0.5cm 0.8cm 0.5cm 1.2cm},clip,width=0.2\textwidth]{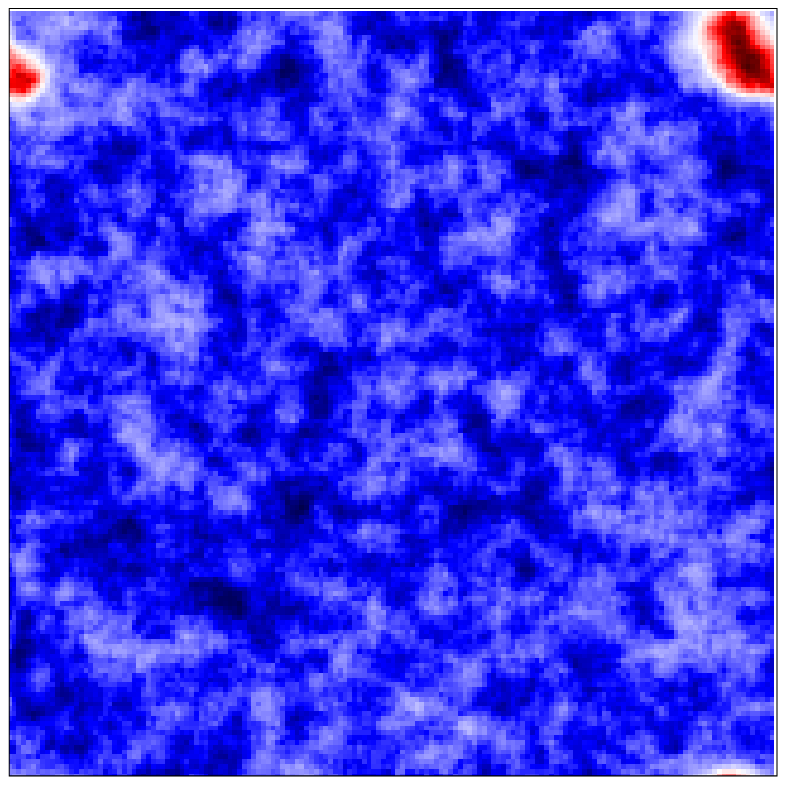}%
  }
  \vspace{-3mm}
  \subfloat[$\phi=\pi$]{%
    \includegraphics[trim={0.5cm 0.8cm 0.5cm 1.2cm},clip,width=0.2\textwidth]{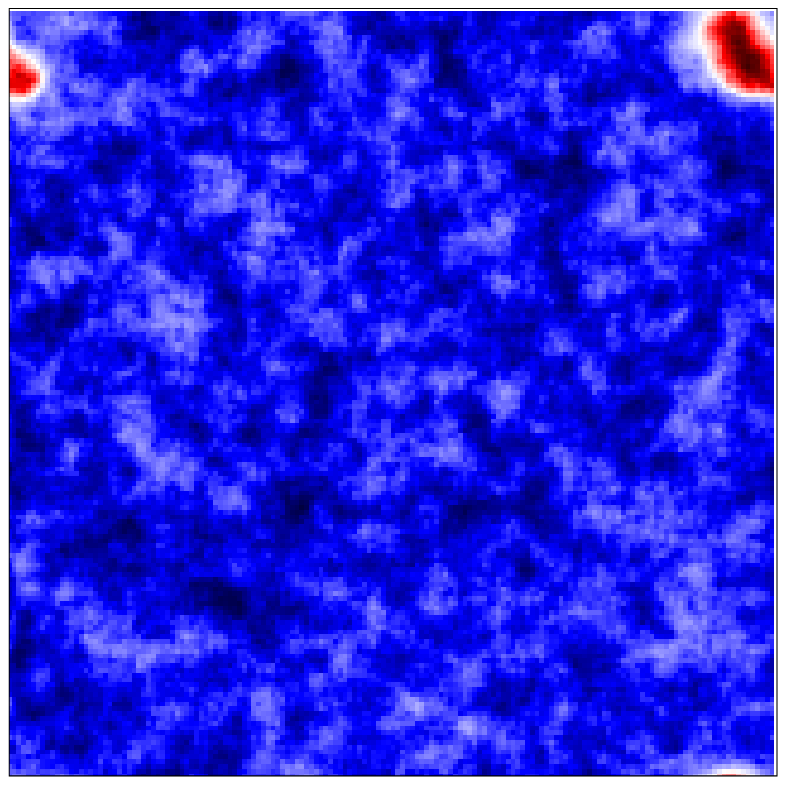}%
  }
  \subfloat[$\phi=1.2\pi$]{%
    \includegraphics[trim={0.5cm 0.8cm 0.5cm 1.2cm},clip,width=0.2\textwidth]{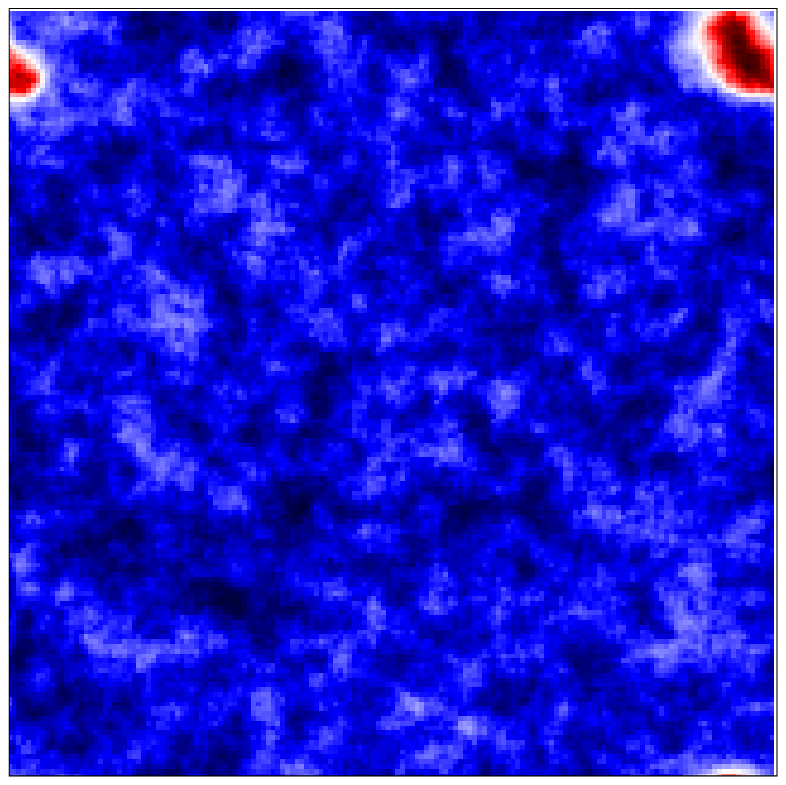}%
  }
  \subfloat[$\phi=1.4\pi$]{%
    \includegraphics[trim={0.5cm 0.8cm 0.5cm 1.2cm},clip,width=0.2\textwidth]{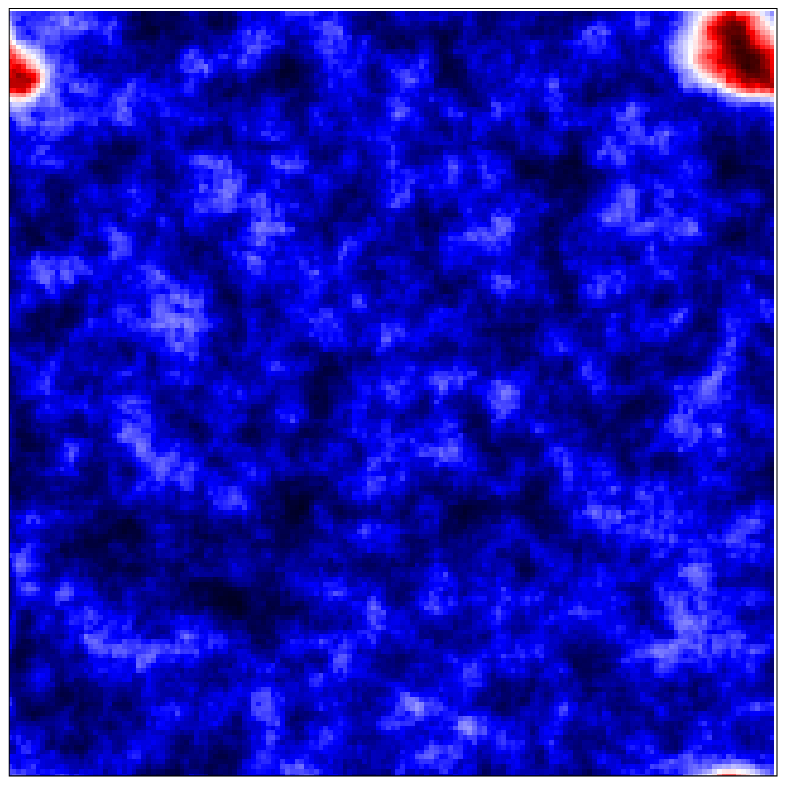}%
  }
  \subfloat[$\phi=1.6\pi$]{%
    \includegraphics[trim={0.5cm 0.8cm 0.5cm 1.2cm},clip,width=0.2\textwidth]{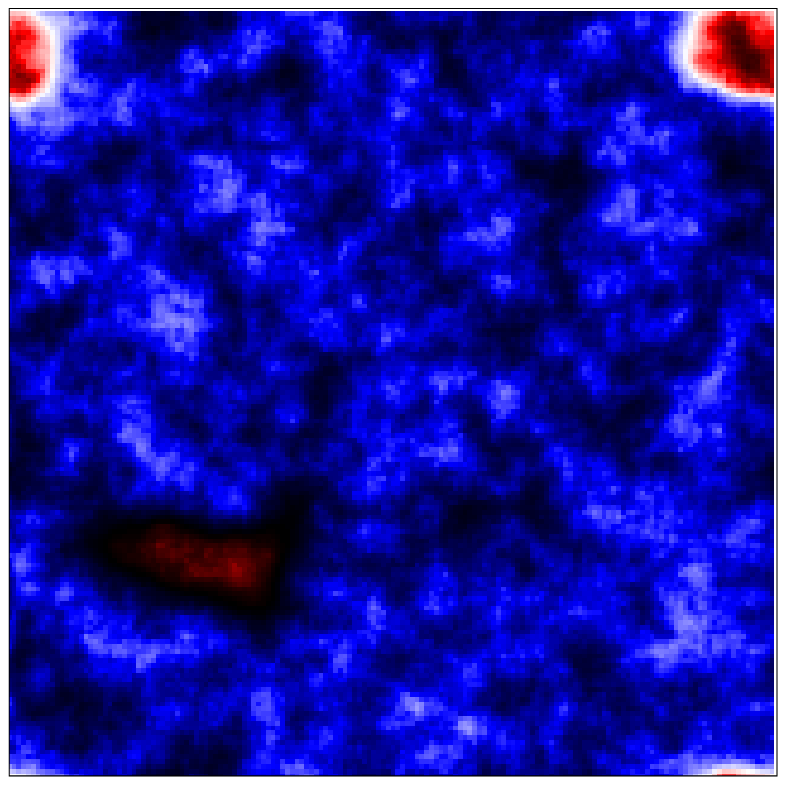}%
  }
  \subfloat[$\phi=1.8\pi+4n\pi$]{%
    \includegraphics[trim={0.5cm 0.8cm 0.5cm 1.2cm},clip,width=0.2\textwidth]{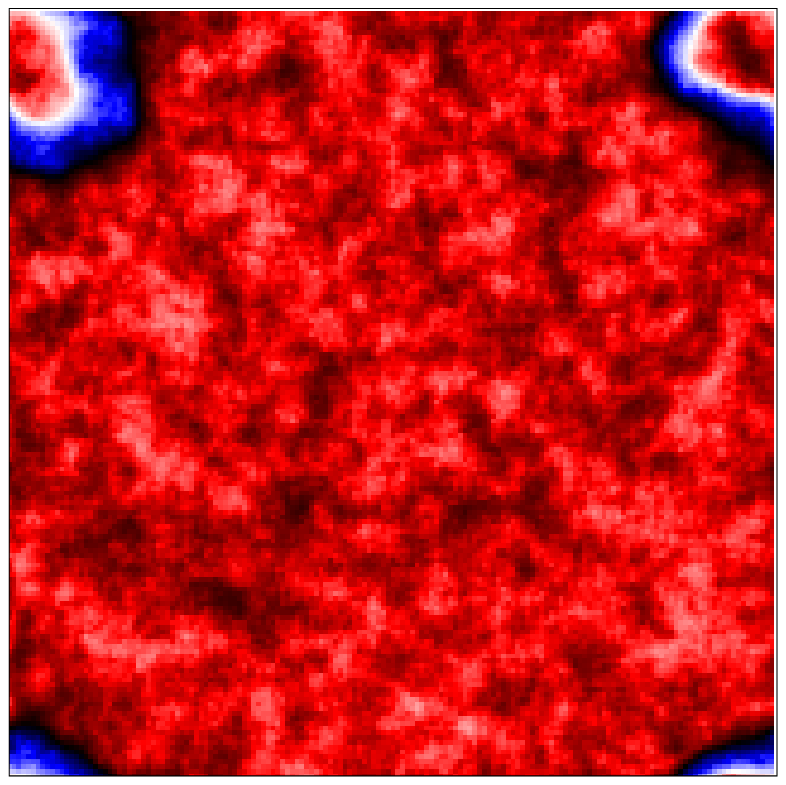}%
  }
  \vspace{-3mm}
  \subfloat[$\phi=2\pi+4n\pi$]{%
    \includegraphics[trim={0.5cm 0.8cm 0.5cm 1.2cm},clip,width=0.2\textwidth]{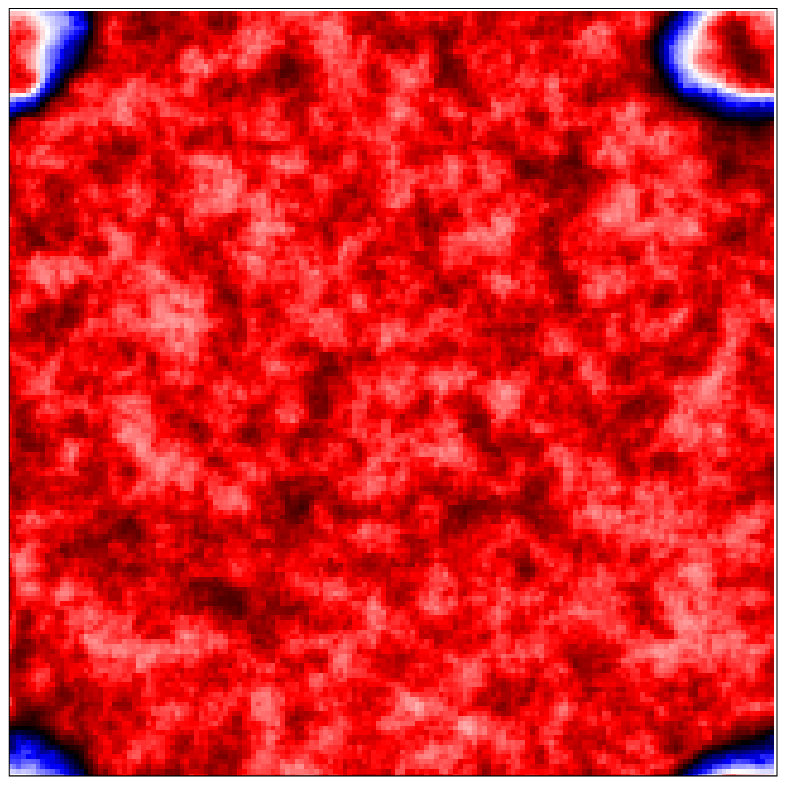}%
  }
  \subfloat[$\phi=2.2\pi+4n\pi$]{%
    \includegraphics[trim={0.5cm 0.8cm 0.5cm 1.2cm},clip,width=0.2\textwidth]{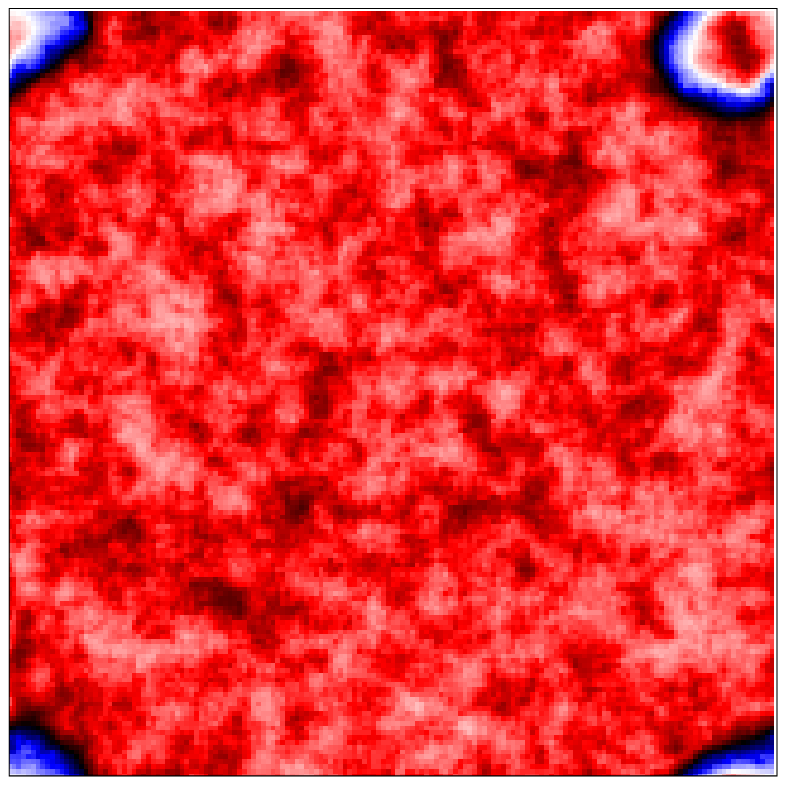}%
  }
  \subfloat[$\phi=2.4\pi+4n\pi$]{%
    \includegraphics[trim={0.5cm 0.8cm 0.5cm 1.2cm},clip,width=0.2\textwidth]{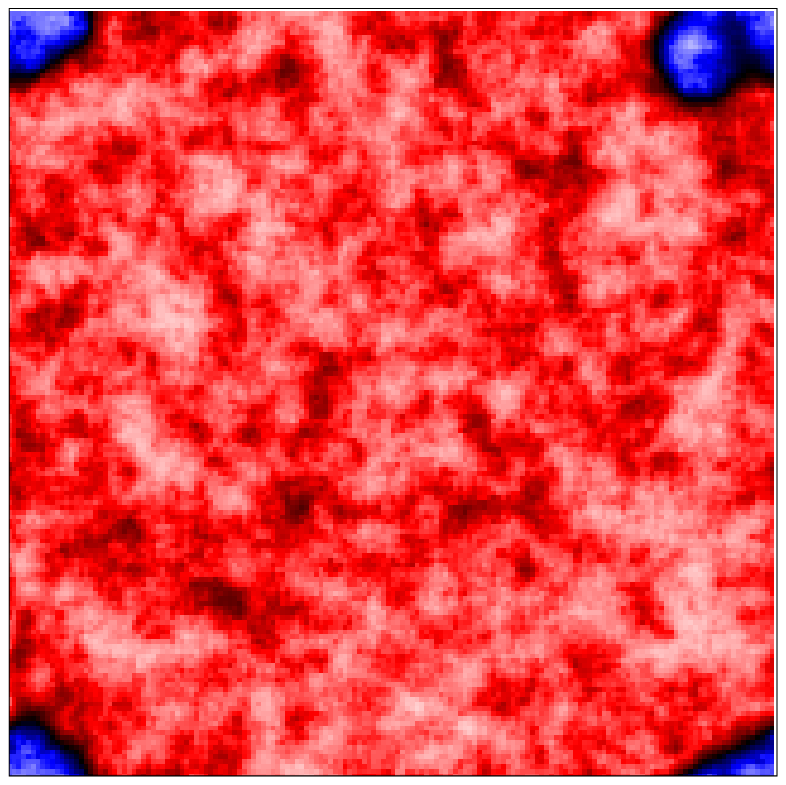}%
  }
  \subfloat[$\phi=2.6\pi+4n\pi$]{%
    \includegraphics[trim={0.5cm 0.8cm 0.5cm 1.2cm},clip,width=0.2\textwidth]{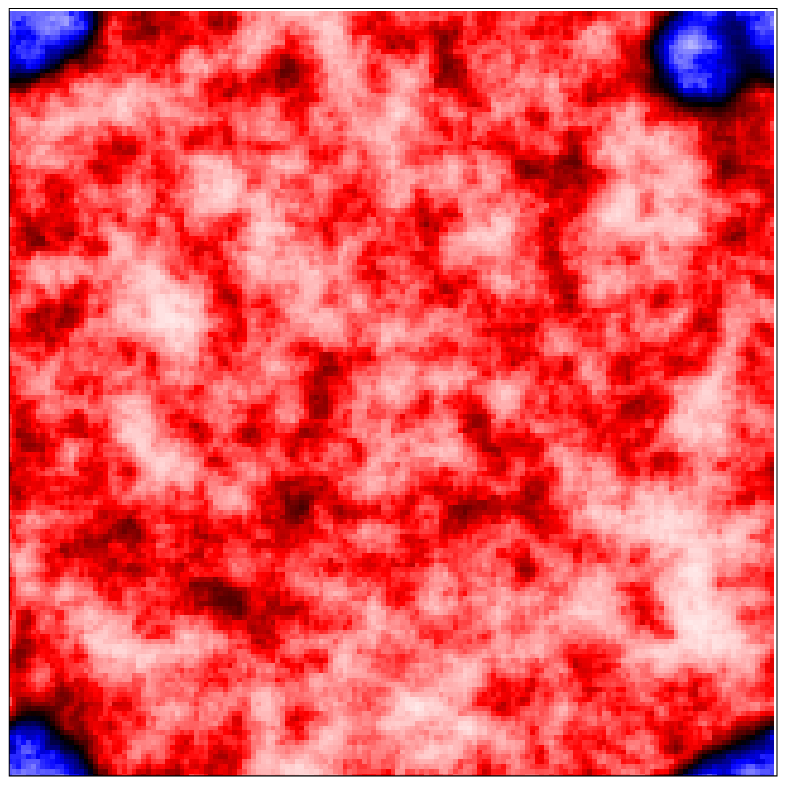}%
  }
  \subfloat[$\phi=2.8\pi+4n\pi$]{%
    \includegraphics[trim={0.5cm 0.8cm 0.5cm 1.2cm},clip,width=0.2\textwidth]{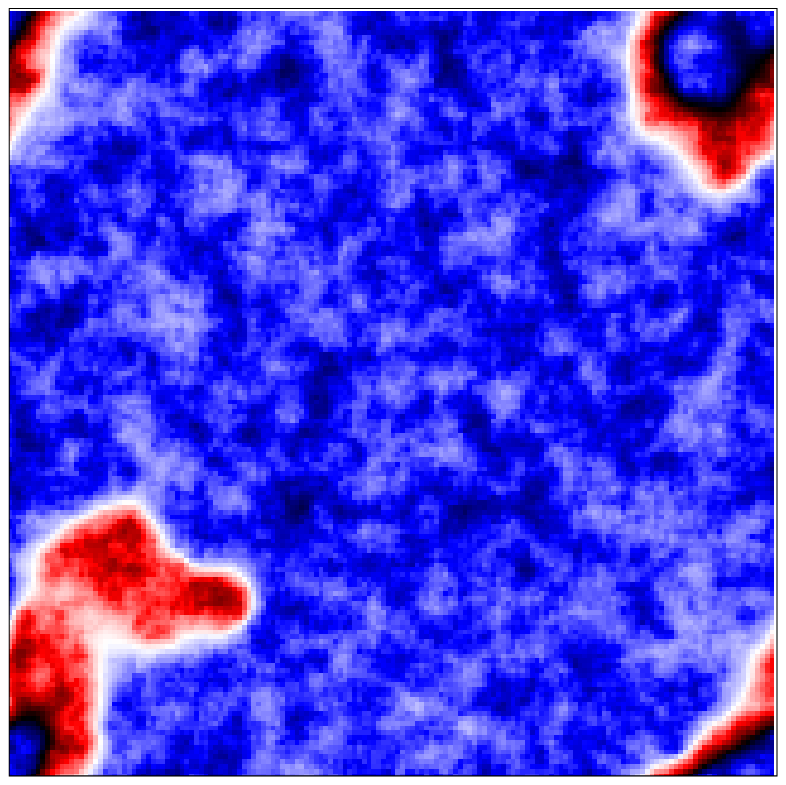}%
  }
  \vspace{-3mm}
  \subfloat[$\phi=3\pi+4n\pi$]{%
    \includegraphics[trim={0.5cm 0.8cm 0.5cm 1.2cm},clip,width=0.2\textwidth]{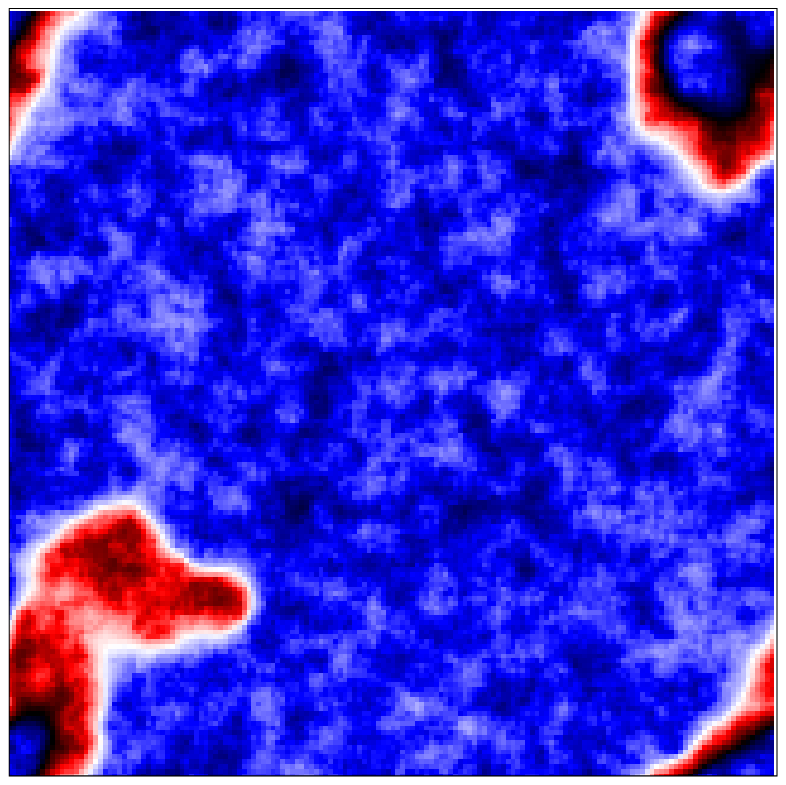}%
  }
  \subfloat[$\phi=3.2\pi+4n\pi$]{%
    \includegraphics[trim={0.5cm 0.8cm 0.5cm 1.2cm},clip,width=0.2\textwidth]{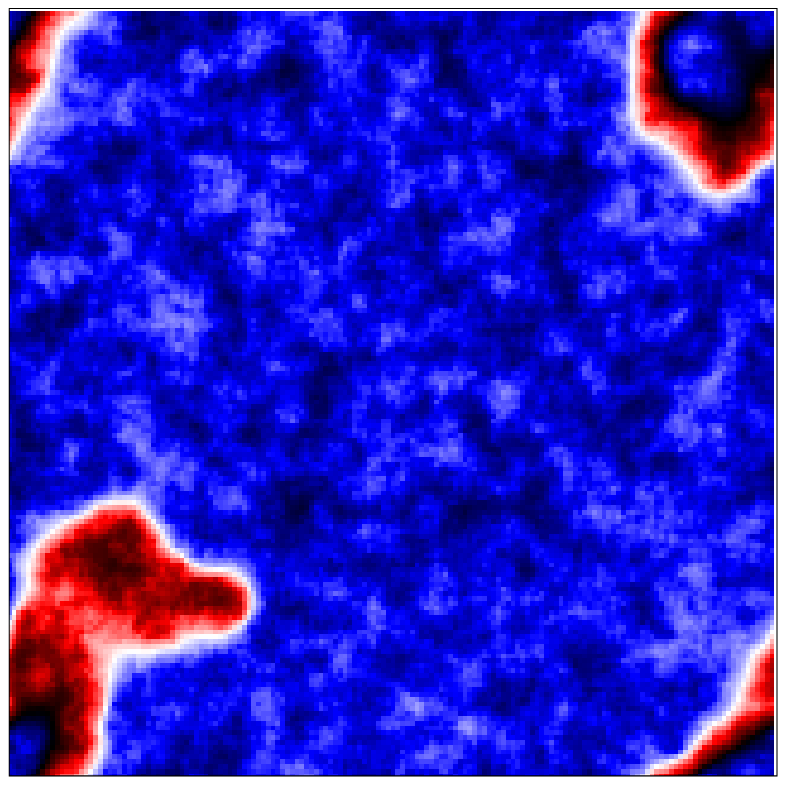}%
  }
  \subfloat[$\phi=3.4\pi+4n\pi$]{%
    \includegraphics[trim={0.5cm 0.8cm 0.5cm 1.2cm},clip,width=0.2\textwidth]{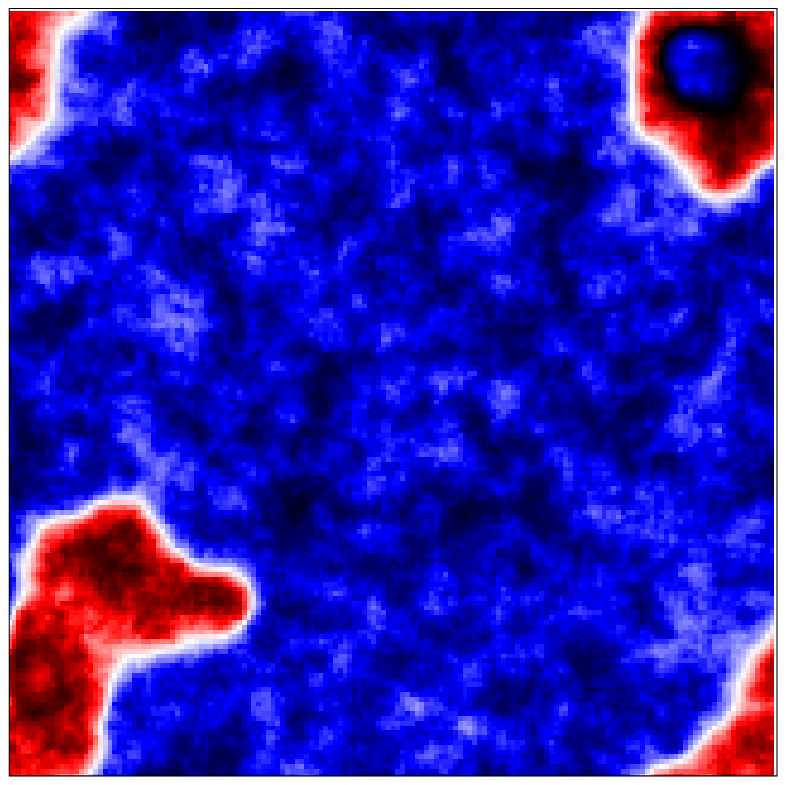}%
  }
  \subfloat[$\phi=3.6\pi+4n\pi$]{%
    \includegraphics[trim={0.5cm 0.8cm 0.5cm 1.2cm},clip,width=0.2\textwidth]{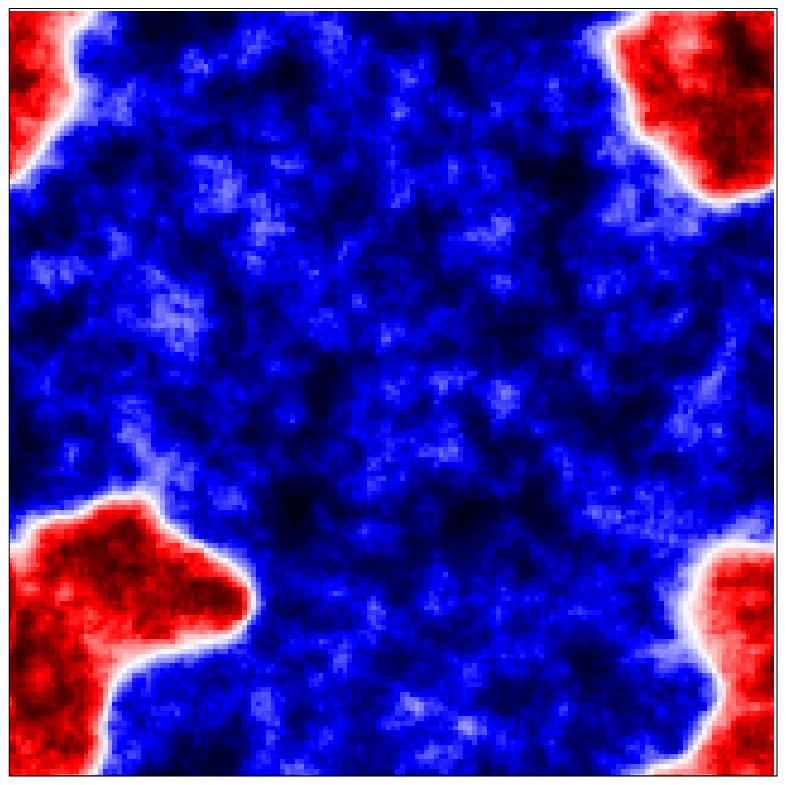}%
  }
  \subfloat[$\phi=3.8\pi+4n\pi$]{%
    \includegraphics[trim={0.5cm 0.8cm 0.5cm 1.2cm},clip,width=0.2\textwidth]{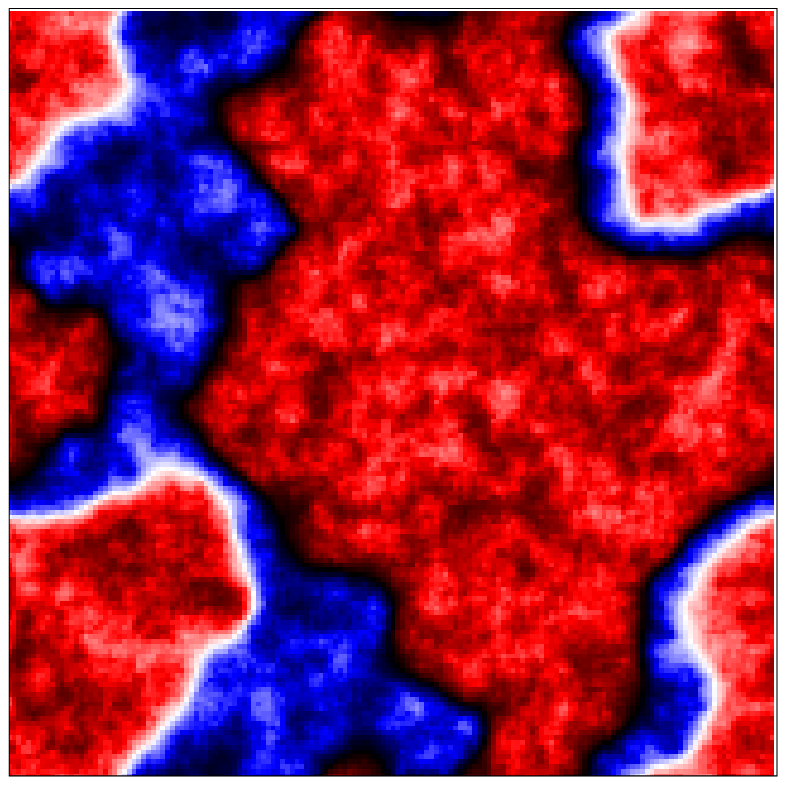}%
  }
  \vspace{-3mm}
  \subfloat[$\phi=4\pi+4n\pi$]{%
    \includegraphics[trim={0.5cm 0.8cm 0.5cm 1.2cm},clip,width=0.2\textwidth]{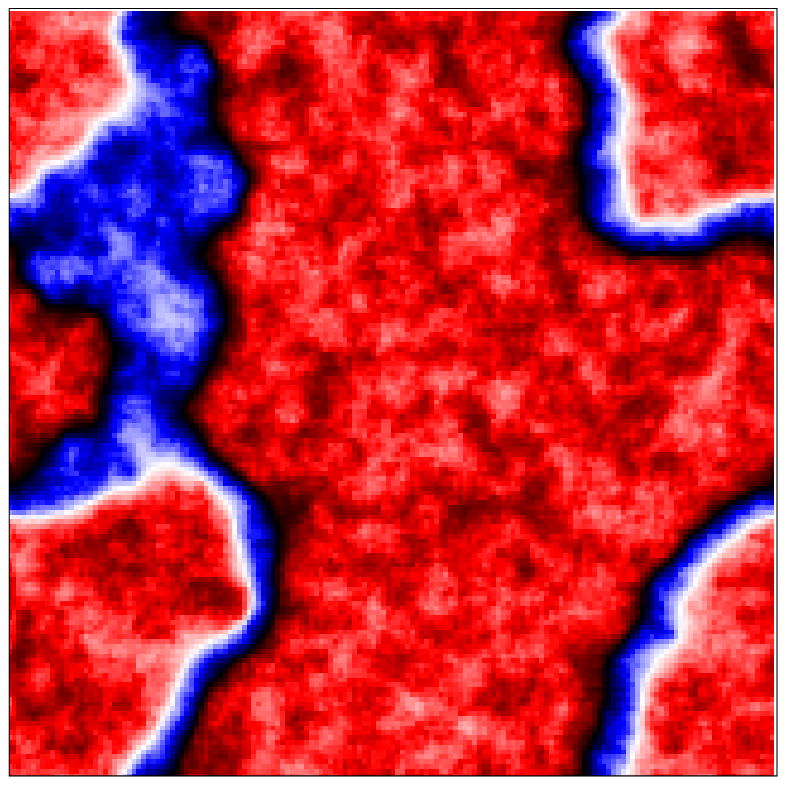}%
  }
  \subfloat[$\phi=4.2\pi+4n\pi$]{%
    \includegraphics[trim={0.5cm 0.8cm 0.5cm 1.2cm},clip,width=0.2\textwidth]{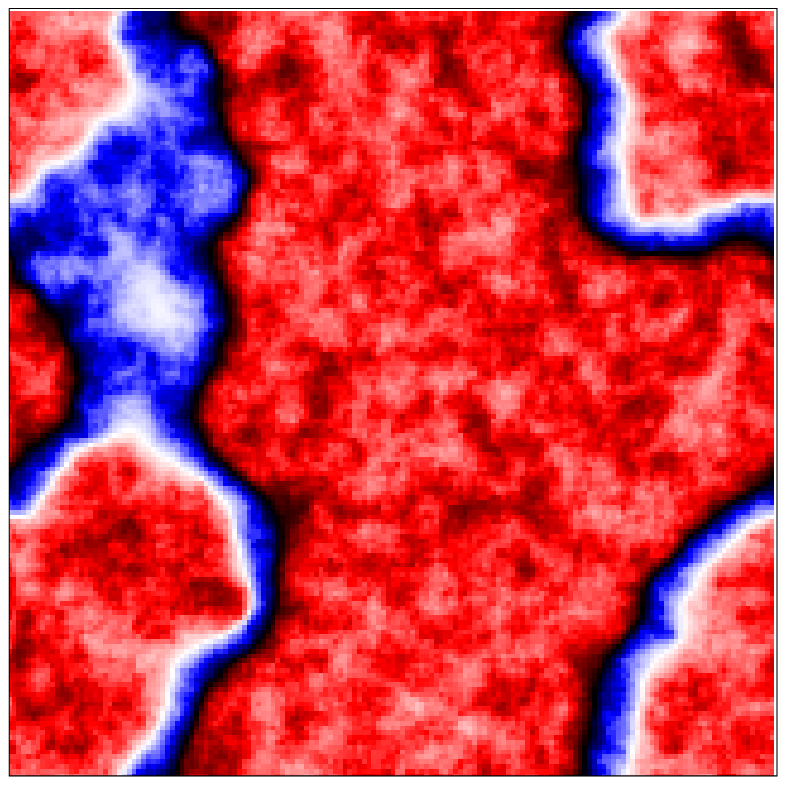}%
  }
  \subfloat[$\phi=4.4\pi+4n\pi$]{%
    \includegraphics[trim={0.5cm 0.8cm 0.5cm 1.2cm},clip,width=0.2\textwidth]{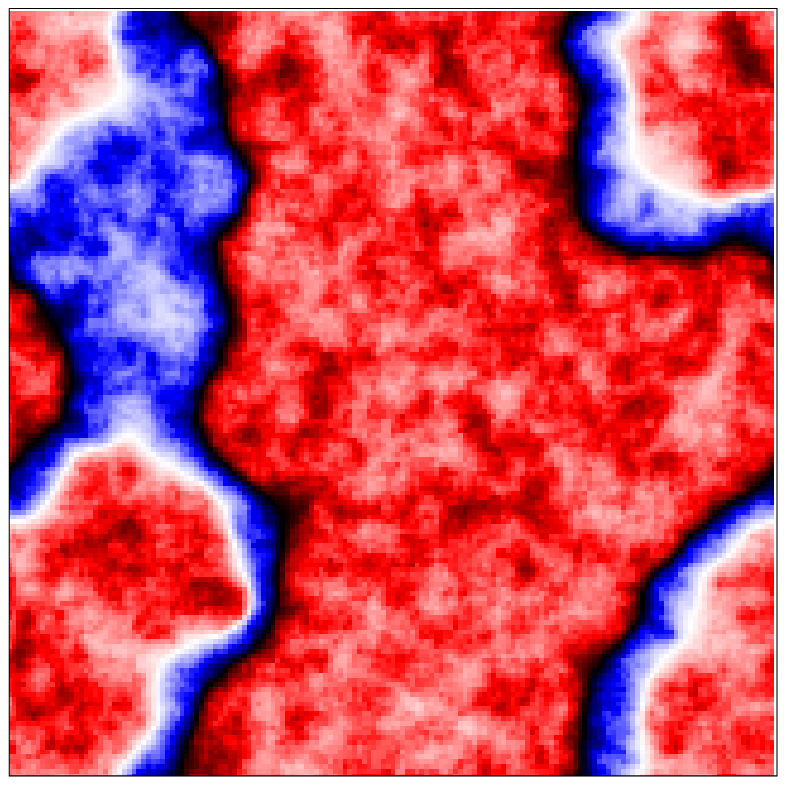}%
  }
  \subfloat[$\phi=4.6\pi+4n\pi$]{%
    \includegraphics[trim={0.5cm 0.8cm 0.5cm 1.2cm},clip,width=0.2\textwidth]{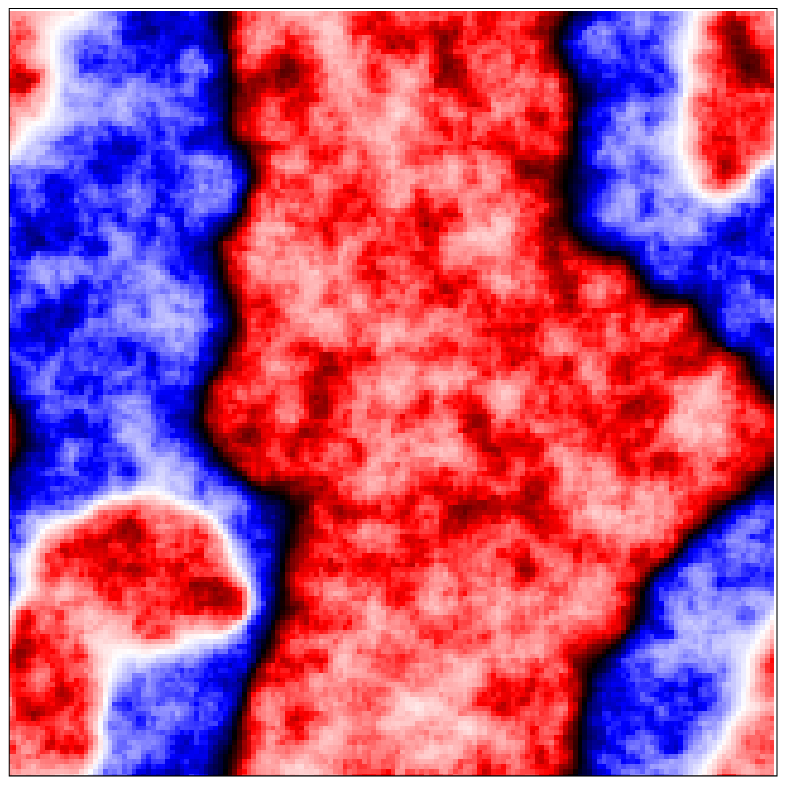}%
  }
  \subfloat[$\phi=4.8\pi+4n\pi$]{%
    \includegraphics[trim={0.5cm 0.8cm 0.5cm 1.2cm},clip,width=0.2\textwidth]{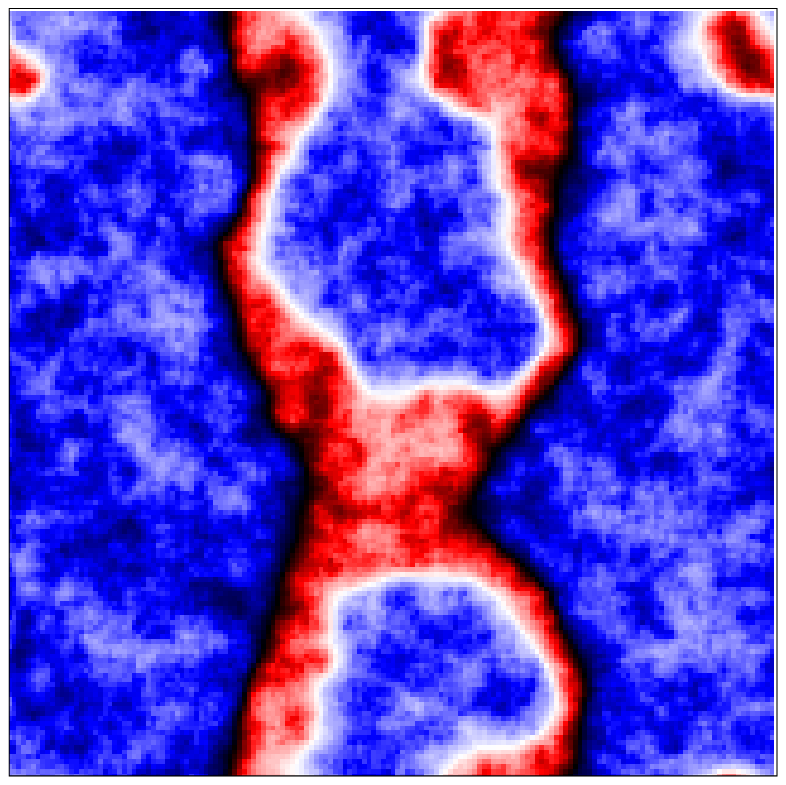}%
  }
  \vspace{-6mm}
  \subfloat[$\phi=5\pi+4n\pi$]{%
    \includegraphics[trim={0.5cm 0.8cm 0.5cm 1.2cm},clip,width=0.2\textwidth]{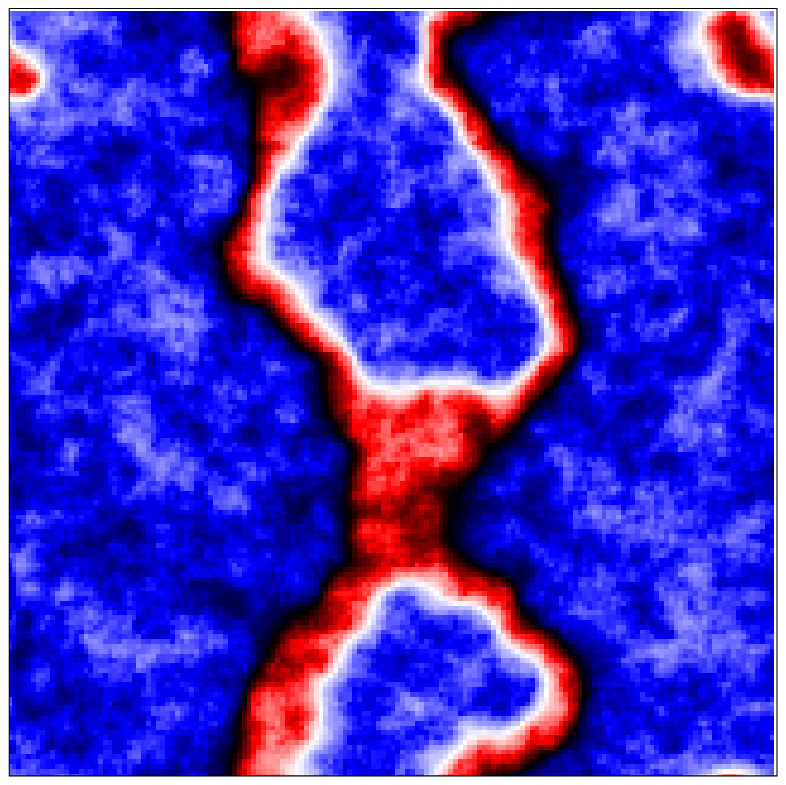}%
  }
  \subfloat[$\phi=5.2\pi+4n\pi$]{%
    \includegraphics[trim={0.5cm 0.8cm 0.5cm 1.2cm},clip,width=0.2\textwidth]{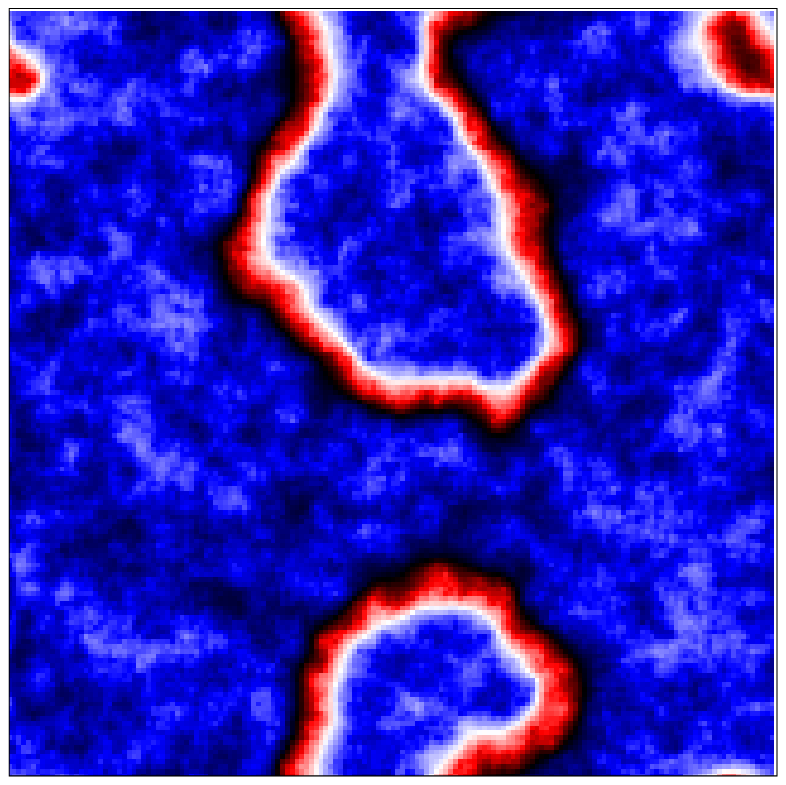}%
  }
  \subfloat[$\phi=5.4\pi+4n\pi$]{%
    \includegraphics[trim={0.5cm 0.8cm 0.5cm 1.2cm},clip,width=0.2\textwidth]{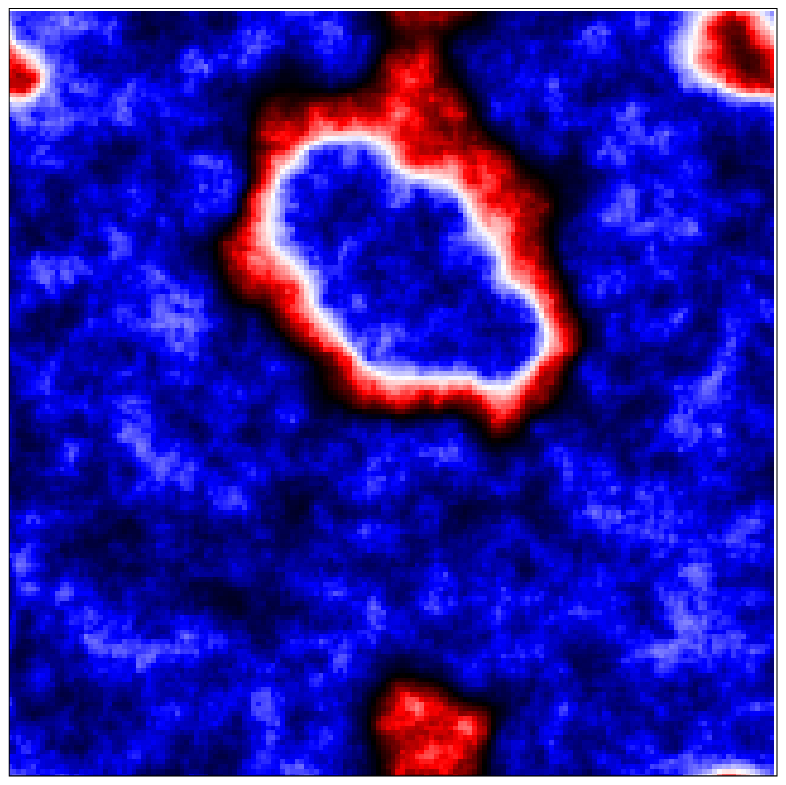}%
  }
  \subfloat[$\phi=5.6\pi+4n\pi$]{%
    \includegraphics[trim={0.5cm 0.8cm 0.5cm 1.2cm},clip,width=0.2\textwidth]{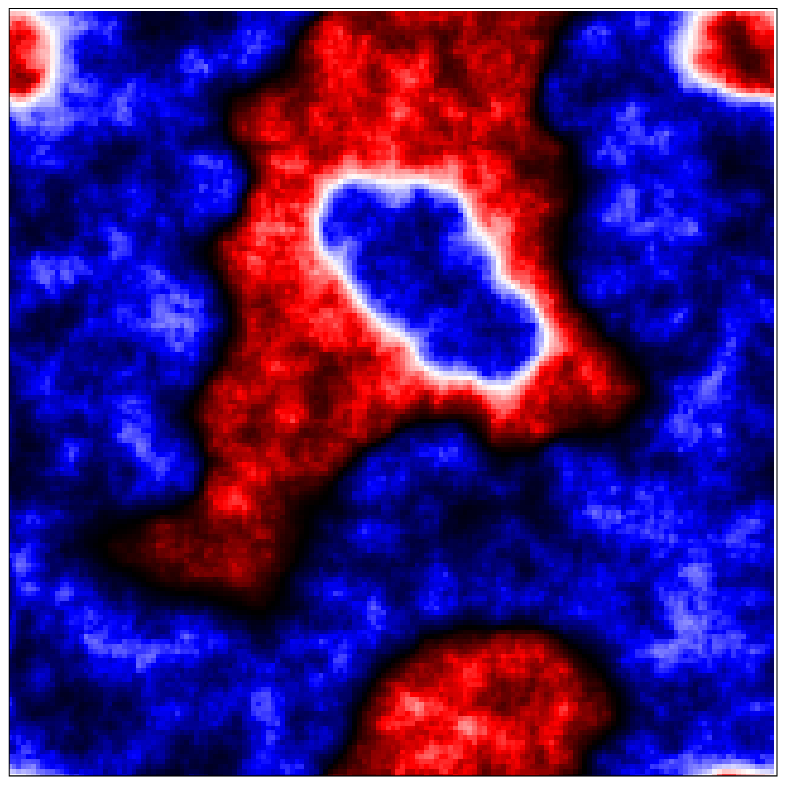}%
  }
  \subfloat[Spin Response]{%
  \resizebox{0.2\textwidth}{!}{%
      \input{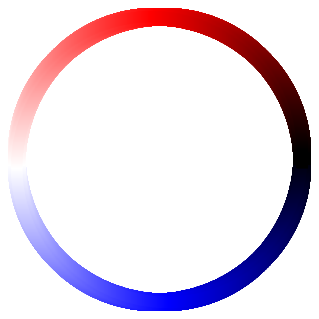}}%
  }
  \caption{
    Example of spin configurations during a period-2 limit cycle with transient response of less than $2\pi$.  
    Spin configurations (a-i) is transient response which does repeat.
    Spin configurations (j-cc) are for limit cycle of $4\pi$  period which is double the periodicity of the driving field angle $\phi$. 
  }
  \label{fig:phase_plots}
  \end{center}
\end{figure*}

Figure \ref{fig:phase_plots} shows the various spin configurations the system goes through before and during a limit cycle with periodicity $4\pi$. The rich structure of the domain walls are stable due to the random field distribution. All the plots in Fig.~\ref{fig:phase_plots} are unique and Figs.~\ref{fig:phase_plots}(a-i) does not repeat but Figs.~\ref{fig:phase_plots}(j-cc) are part of the limit cycle which repeats indefinitely. See \cite{PURR3260} for better visualizations of the evolving spin configurations in limit cycles with $n>1$ periods.

\subsection{Response with fluctuations in temperature}
\label{ssxn:Response_with_fluctuations_in_temperature}

\begin{figure*}[htbp]
  \begin{center}
    \subfloat[ $H = 0.058 J ;~ N = 64 \times 64. $\label{sfig:temp_perturb_T0.1_L64_n2}]{%
      \resizebox{.3\textwidth}{!}{%
        \input{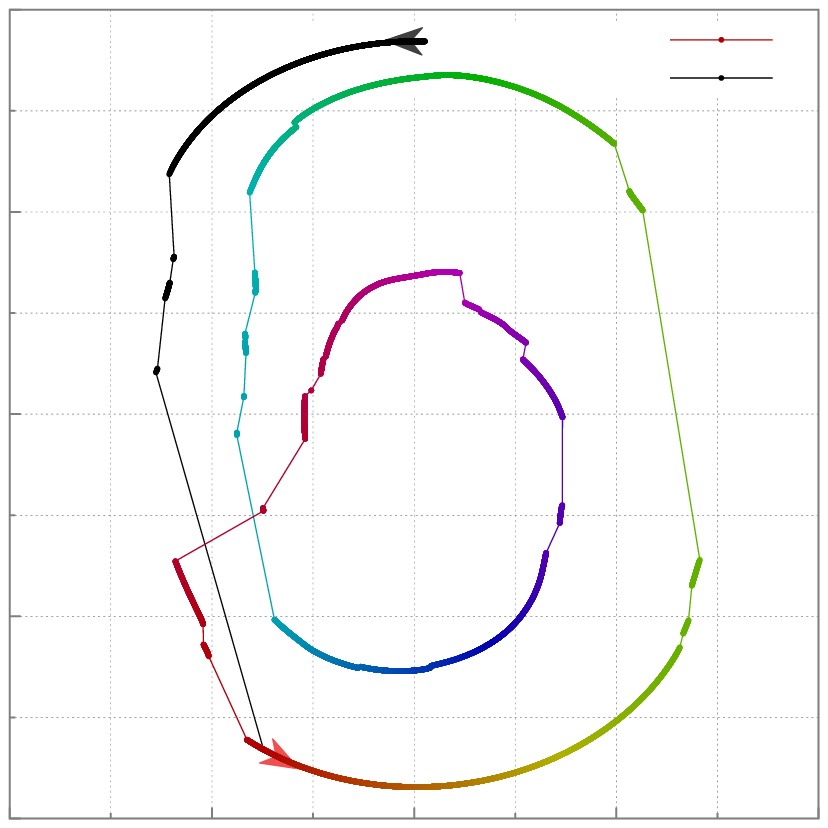}}%
    }%
    \subfloat[ $H = 0.052 J ;~ N = 100 \times 100. $\label{sfig:temp_perturb_T0.1_L100_n3}]{%
      \resizebox{.3\textwidth}{!}{%
        \input{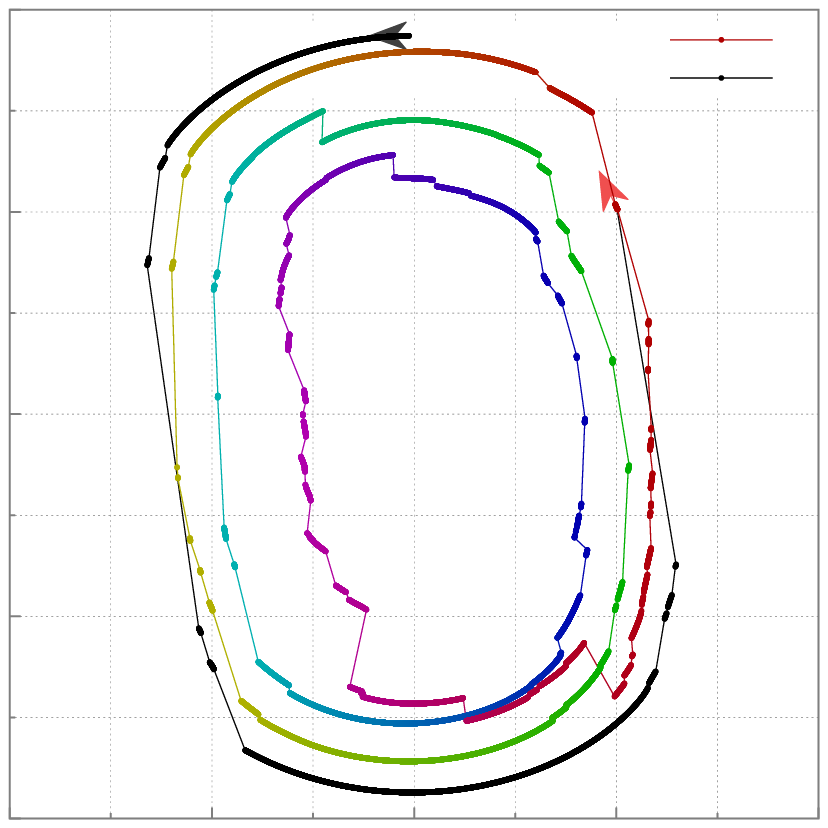}}%
    }%
    \subfloat[ $H = 0.045 J ;~ N = 128 \times 128. $\label{sfig:temp_perturb_T0.1_L128_n4}]{%
      \resizebox{.3\textwidth}{!}{%
        \input{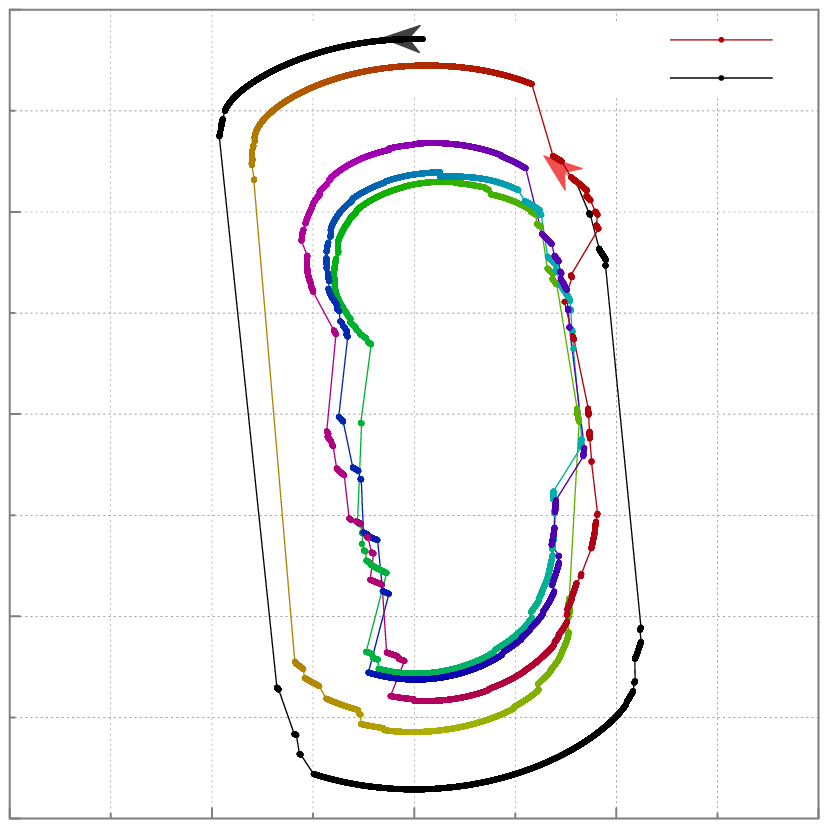}}%
    }%
    \\
    \subfloat[ $ H = 0.044 J ;~ N = 160 \times 160. $\label{sfig:temp_perturb_T0.1_L160_n5}]{%
      \resizebox{.45\textwidth}{!}{%
        \input{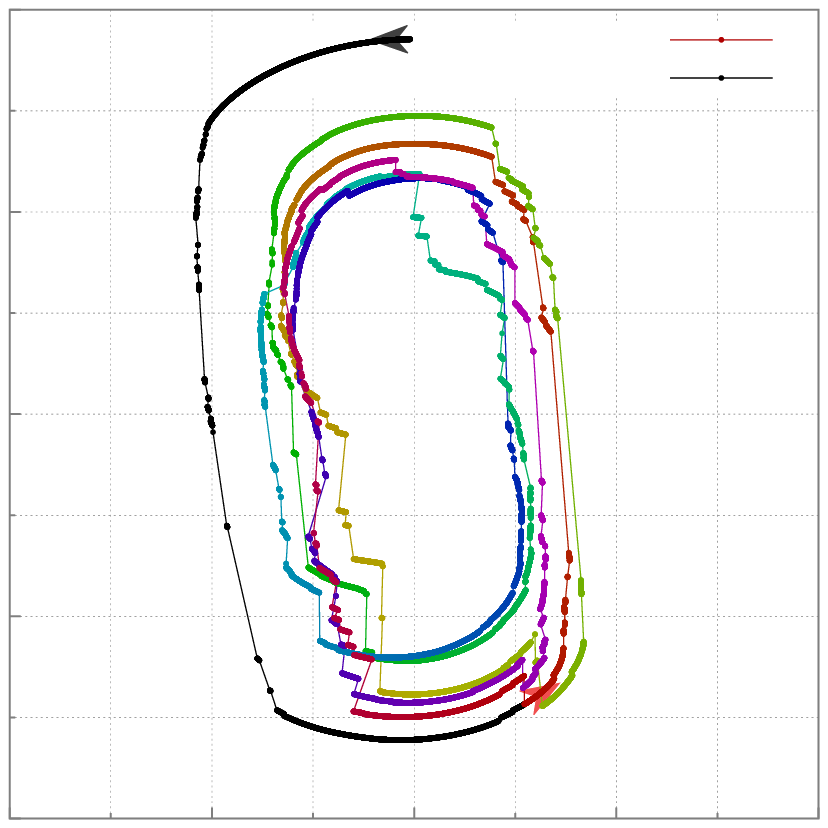}}%
    }%
    \subfloat[ $ H = 0.048 J ;~ N = 160 \times 160. $\label{sfig:temp_perturb_T0.1_L160_n7}]{%
      \resizebox{.45\textwidth}{!}{%
        \input{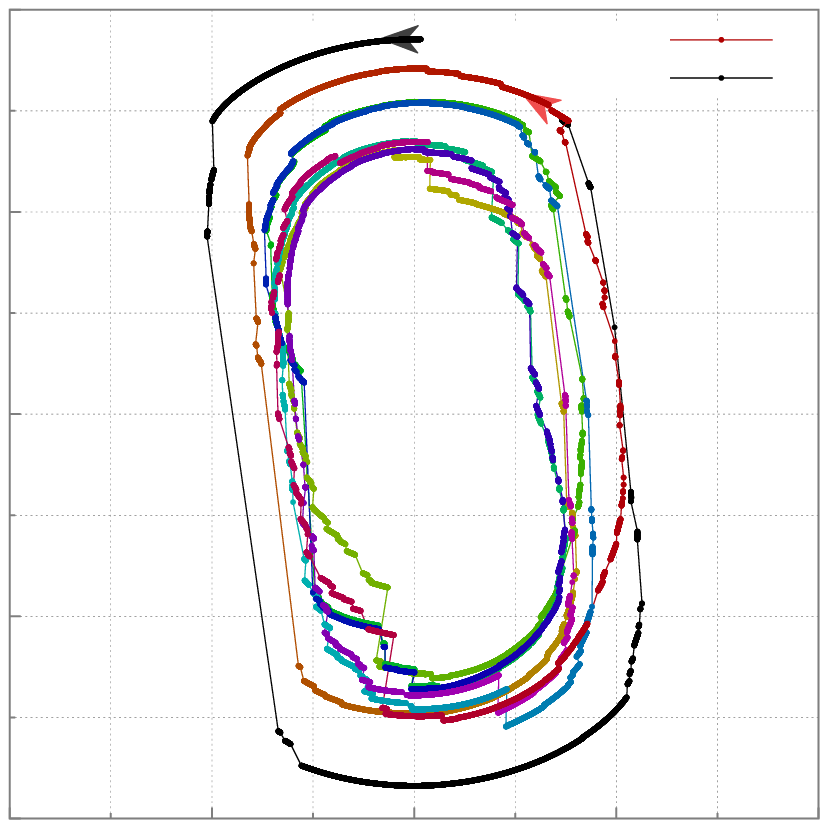}}%
    }%
    \caption{\textcolor{black}{Transient response and multiperiod limit cycles near the transition with finite temperature fluctuation. These are results from simulations with  the protcol decribed in Sec.~\ref{ssxn:Response_with_fluctuations_in_temperature} at T=0.1J for the Monte Carlo sweeps in between spin relaxation steps.}
    Panels (a-e) show the initial transient response (black curves), followed by
    multiperiodic limit cycles (rainbow curves). These cycles are repeated for several periods and all of them stay in the same limit cycle as its zero-temperature counterpart till the end of the simulations. 
    (a) Here, the period-2 limit cycle continues for $\Delta\phi=375*2\pi$
    (b) Here, the period-3 limit cycle continues for $\Delta\phi=300*2\pi$
    (c) Here, the period-4 limit cycle continues for $\Delta\phi=250*2\pi$
    (d) Here, the period-5 limit cycle continues for $\Delta\phi=221*2\pi$
    (e) Here, the period-7 limit cycle continues for $\Delta\phi=230*2\pi$
    }
  \label{fig:temp_perturb}
  \end{center}
\end{figure*}

\textcolor{black}{
In order for a physical system to remain at very low temperature
under the influence of a driving field, it is necessary for it 
to be connected to a low temperature heat bath which 
carries away the heat generated by the driving field in an efficient manner.
This implies that the physical system will experience temperature
fluctuations which arise from the heat bath.  
}
We model these temperature fluctuations with a Monte Carlo sweep in between spin relaxation.
\textcolor{black}{
As long as fluctuations which arise due to temperature are slow compared
to the relaxation times of the system, this is a reasonable model for very low temperatures.
}
We find (Fig.~\ref{fig:temp_perturb}) that the multiperiod limit cycles are still rigid against the
\textcolor{black}{low}
temperature fluctuations which are simulated by the following protocol:
\begin{enumerate}
  \item Initialize spins to fully ordered in the $+y$ direction. Set $\phi=\pi/2$.
  \item After updating $\phi$:
  \begin{enumerate}
    \item Update spins using the spin relaxation method described in the main text (Sec-IV).
    \item One Monte-Carlo sweep at temperature $(T=0.1J)$ over the whole lattice using Glauber dynamics with checkerboard updates (This comprises one checkerboard update over all the black sites followed by all the white sites).
    \item Update spins using the spin relaxation method again.
  \end{enumerate}
  \item Update $\phi\rightarrow\phi+\delta\phi(=0.0001*2\pi)$. Go to step-2.
\end{enumerate}

This implies that the classical discrete Time Crystal(CDTC) we found is robust against small temperature fluctuations. 
Our CDTC is also interesting due to the fact that it has only short range interactions.

\subsection{Gaussian Random Fields}
For uniaxial random field disorder in the $x$ direction,
a local random field $h_{x,i} $ is chosen at each site $i$ 
from a Gaussian distribution:
\begin{eqnarray}
  P(h_{x,i})=  \frac{1}{\sqrt{2\pi R_{x}^2}} \exp(-\frac{h_{x,i}^2}{2 R_{x}^2})
  \label{eqn:gaussian_disorder}
\end{eqnarray}

Because this is an unbounded probability distribution, the question arises as
to whether we have accurately captured the behavior of the system 
in the presence of ``rare events''.  
To quantify the likelihood of a rare event, 
we ask the question:  how large of a system size $N = L \times L$
do we need to simulate in order to have at least a 
$1 \%$ chance that an event as rare as $|h_i| > 5 R_x$  happens?
The answer is a system of size at least $N \gtrsim 132 \times 132$.

This can be seen as follows. 
The probability that there is at least one site $i \in N$ for which 
 $|h_i| > aR $ 
\begin{equation} 
P(\exists i \in N~{\rm s.t.}~|h_i| > a R)
\end{equation}
is equal to the complement of the probability that 
$ |h_i| \le aR,\ \forall i $:
\begin{equation}
P(\exists i \in N~{\rm s.t.}~|h_i| > a R) = 1 - P(|h_i| \le a R,\ \forall i), \\ 
\end{equation}
and
\begin{eqnarray}
P(|h_i| \le a R,\ \forall i)  &=& [P(|h_i| \le a R)]^N \nonumber \\
&=& [\textrm{erf}(a/\sqrt{2})]^N
\end{eqnarray}
where
\begin{equation}
\textrm{erf}(x) = \int_{-x}^{x} \frac{1}{\sqrt{\pi}}\exp(-y^2) dy
\end{equation}
is the error function.  

Then the required system size to have a 1\%  chance for such an event to occur is given by: 
\begin{eqnarray}
N&=&L^d \nonumber \\
&=&\frac{\log(1-[P(\exists i \in N {\rm s.t.} |h_i| > a R)\equiv0.01])}{\log(\textrm{erf}(a/\sqrt{2}))}
\end{eqnarray}
With $a = 5$ and $ d=2$,
we find that  $ L \ge 132$.


\subsection{Equilibrium Results}
\label{sxn:Field_Cooling_vs_Zero_field_Cooling}


\begin{figure*}[htbp]
  \begin{center}
  \subfloat[ Susceptibility of y-magnetization at $ R_x=0.5 J$]{%
  \label{sfig:RFXY_unax_FC_Xyy}%
  \resizebox{.5\textwidth}{!}{%
      \input{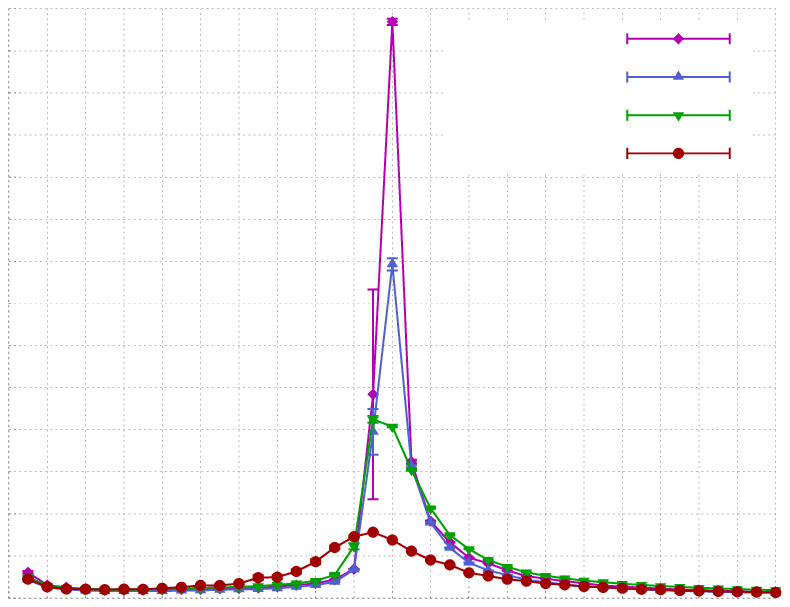}}%
  }
  \subfloat[ Binder parameter {\em vs.} temperature, at $ R_x=0.5 J$ ]{%
  \label{sfig:RFXY_unax_FC_B}%
  \resizebox{.5\textwidth}{!}{%
      \input{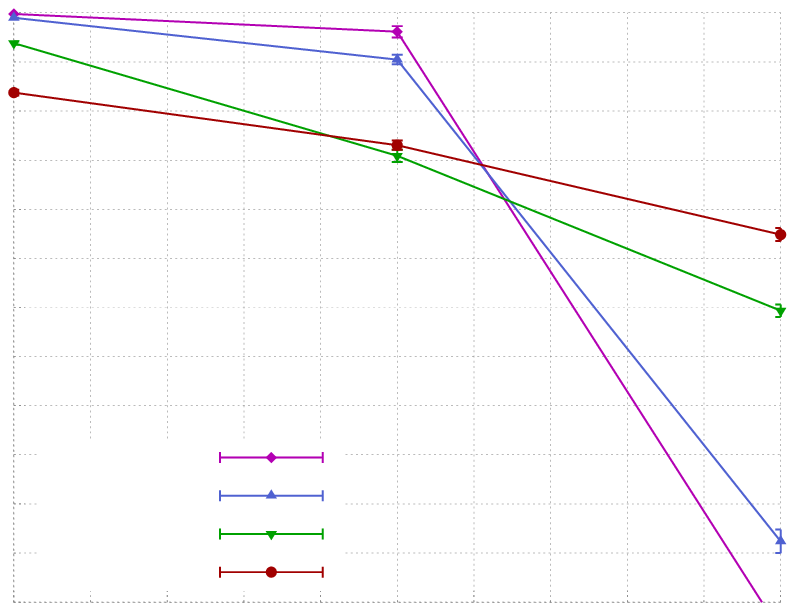}}%
  }
  \caption{Susceptibility to order and Binder parameter at moderate uniaxial disorder strength, $R_x = 0.5J$.  
    (a) The magnetic susceptibility  ($\chi_{yy}$)  in the $y$ direction 
peaks near  $ T_c\simeq J $, and diverges 
as system size is increased. 
    (b) The Binder parameter yields a transition temperature $ T_c\simeq 0.96J $, consistent with
    the peak in the magnetic susceptibility shown in panel (a).
  }
  \label{fig:RFXY_unax_FC_Tc}
  \end{center}
\end{figure*}

\begin{figure*}[htbp]
  \begin{center}
  \subfloat[ $R_x=0.5, R_y=0.0, m_x(T),\ N=128^2$]{%
  \label{sfig:RFXY_unax_FC_mx}%
  \resizebox{.5\textwidth}{!}{%
    \input{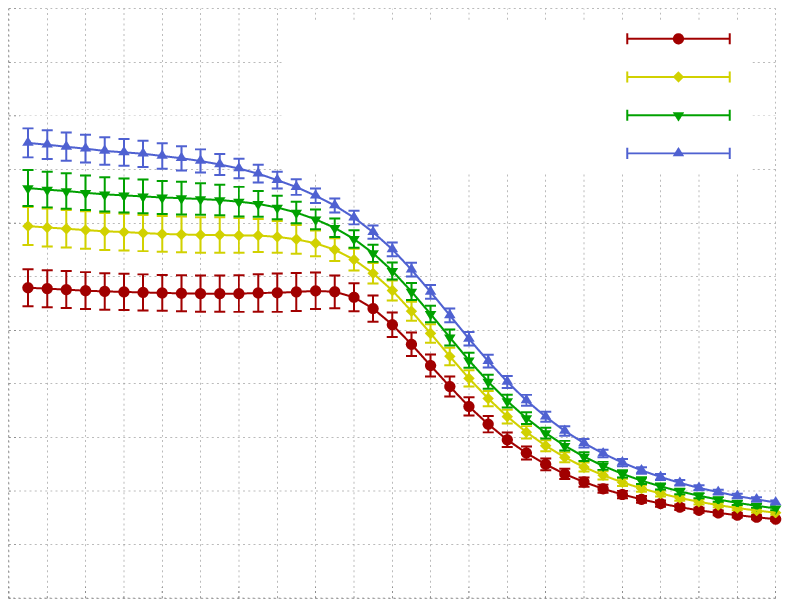}}%
  }
  \subfloat[ $R_x=0.5, R_y=0.0, m_y(T),\ N=128^2$]{%
  \label{sfig:RFXY_unax_FC_my}%
  \resizebox{.5\textwidth}{!}{%
    \input{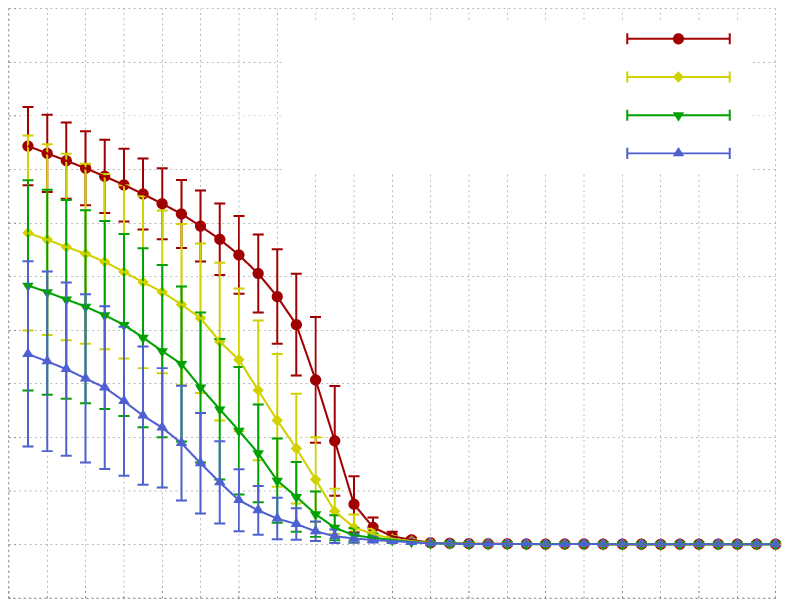}}%
  }
  \caption{Transverse field cooling at $R_x = 0.5 J$.  
(a) Magnetization in the $x$ direction $m_x$   and (b)  magnetization in 
the $y$  direction $m_y$ in the presence of
both uniaxial random field disorder $R_x$ 
{\em and} an applied uniform field $H_x$.
The spontaneous magnetization $m_y$ 
remains robust at finite disorder strength and in the presence of 
a uniform field applied transverse to the ordering direction.
   }
  \label{fig:RFXY_unax_FC_m}
  \end{center}
\end{figure*}

In this section, we report our results 
from Monte Carlo simulations of 
Eqn.~1 in the main text  
in thermal 
equilibrium.
We employ a Metropolis algorithm with checkerboard updates, in which one Monte-Carlo sweep (MCS)
updates black sites and then white sites.
We follow a field-cooling protocol in which the system is started at 
high temperature of $T=2J$, then we reduce the temperature 
in steps of $\Delta T = 0.05J$ until  $T=0.05J$. 
At each temperature step, we thermalize the system with 128,000 MCS and then take 12,800 measurements which are taken randomly between 1 MCS and 16 MCS.

It is known that the presence of uniaxial random field disorder in the $x$ direction 
($R_x > 0$)
favors spontaneous symmetry breaking in the form of ferromagnetic order in the $y$ direction,\cite{PhysRevB.32.3081,Feldman1999,PhysRevB.74.224448,Crawford2013} via an order-by-disorder mechanism.  
\textcolor{black}{
Bera {\em et al.} have used mean-field theory on the classical XY magnet
to argue that the order-by-disorder phenomenon is robust against applied uniform magnetic field.~\cite{bera_classical}
}
Indeed, our simulations at moderate uniaxial random field strength $R_x = 0.5J$ 
are consistent with spontaneous symmetry breaking in the $y$ direction,
and indicate that this phase is rather robust against disorder strength. 
In Fig.~\ref{fig:RFXY_unax_FC_Tc}, we show that the magnetic susceptibility
in the $y$ direction diverges with system size at the transition temperature
$T_c = 0.96 J$ determined from the Binder parameter.

This order-by-disorder transition is robust even against uniform field applied parallel to the uniaxial random field.
Our simulations of cooling in uniform applied field parallel the uniaxial random field direction
(see Fig.~\ref{fig:RFXY_unax_FC_m}) 
show that an order parameter develops in the direction perpendicular to the 
uniaxial random field, even in the presence of an applied field.
This shows that the spontaneous magnetization $m_y$
is robust even for moderate random field $R_x = 0.5J$, and 
finite uniform applied field $H_x$, as shown in Fig.~\ref{fig:RFXY_unax_FC_m}(b),
consistent with the mean field results of Ref.~\cite{bera_classical}.


\begin{figure}[htbp]
  \begin{center}
  \subfloat[ $T=0.05J,\ R_x=0.5 J,\ R_y=0$]{%
  \label{sfig:RFXY_unax_FC_0.05_L}%
  \resizebox{.5\textwidth}{!}{%
      \input{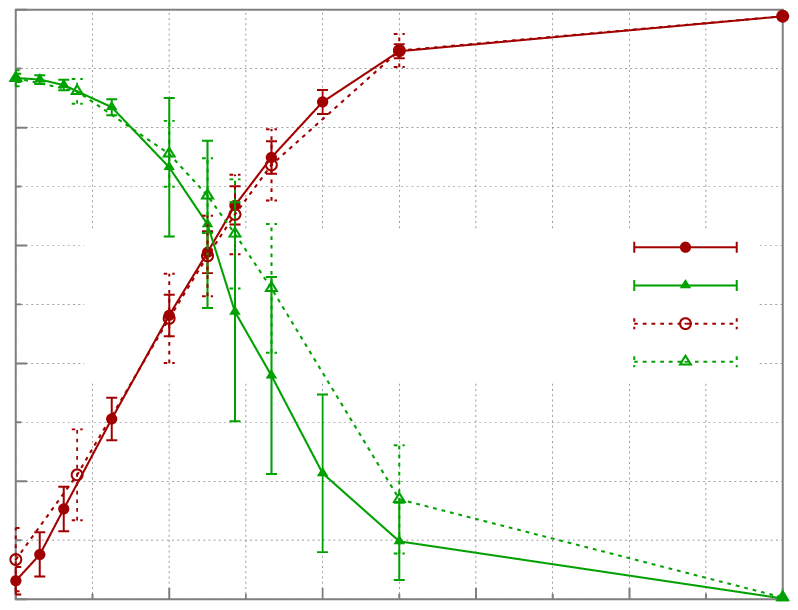}}%
  }
  \caption{Field cooling.  Equilibrium, field-cooled magnetizations in the $x$ and $y$ direction,
    with applied field along the axis of the random field disorder $\vec{H} \parallel R_x$ with $R_x = 0.5J$,
    as described in the text. 
    The horizontal axis is the value of the applied uniform field $H_x$ during the field-cooling protocol. 
Upon field cooling with $H_x \lesssim R_x/10$,   the net magnetization in the y direction $m_y$ dominates over 
    the net magnetization in the x direction $m_x$.  
                This illustrates the robustness of the spontaneous magnetization in the $y$ direction even in the presence of an applied transverse field.  
  }
  \label{fig:RFXY_unax_FC_cross_mx_my}
  \end{center}
\end{figure}


With strong enough transverse applied field $H_x$, 
the order-by-disorder phenomenon must be suppressed
and the system will remain in the paramagnetic phase.  
Fig.~\ref{fig:RFXY_unax_FC_cross_mx_my} shows this crossover of the dominant magnetization from the 
$y$-axis to the $x$-axis with increasing applied transverse field.

\bibliography{RFXY}